\RequirePackage{ifpdf}
\ifpdf
\else
\PassOptionsToPackage{dvipdfmx}{graphicx}
\PassOptionsToPackage{dvipdfmx}{hyperref}
\fi

\documentclass[12pt,a4paper]{article}
\usepackage{jheppub}
\allowdisplaybreaks[4]
\usepackage{amsmath,bm}
\usepackage{subfigure}
\usepackage{slashed}
\usepackage{mathrsfs}

\usepackage{color}
\usepackage[normalem]{ulem}
\usepackage{mathdots}

\newcommand{\pmat}[1]{\begin{pmatrix} #1 \end{pmatrix}}

\newcommand{\nord}[1]{:\mathrel{#1}:}

\newcommand{\Cliff}{\varGamma} 
\renewcommand{\Re}{\operatorname{Re}}
\renewcommand{\Im}{\operatorname{Im}}

\title{
Simulating Quantum Field Theories with Boundaries in Curved Spacetimes Using Open Spin Systems
}

 \author[1]{Shunichiro Kinoshita}
 \author[2]{Keiju Murata}
 \author[2]{Daisuke Yamamoto}
 \author[3,4]{Ryosuke Yoshii}
 \affiliation[1]{Mathematics, Science, Data Science and AI Program, Kanazawa Institute of Technology, Nonoichi, Ishikawa 921-8501, Japan}
 \affiliation[2]{Department of Physics, College of Humanities and Sciences, Nihon University, Sakura-josui,
 Tokyo 156-8550, Japan}
 \affiliation[3]{Center for Liberal Arts and Sciences, Sanyo-Onoda City University, Yamaguchi 756-0884,
 Japan}
 \affiliation[4]{International Institute for Sustainability with Knotted Chiral Meta Matter (WPI-SKCM2), Hiroshima University, Higashi-Hiroshima, Hiroshima 739-8526, Japan}
 \emailAdd{kinoshita@neptune.kanazawa-it.ac.jp}
 \emailAdd{murata.keiju@nihon-u.ac.jp}
 \emailAdd{yamamoto.daisuke21@nihon-u.ac.jp}
 \emailAdd{ryoshii@rs.socu.ac.jp}

\abstract{%
We develop a framework to simulate quantum field theories (QFTs) with boundaries in $(1+1)$-dimenmsional curved spacetimes by employing open spin systems. Building upon our previous work~\cite{Kinoshita:2024ahu} that established a mapping from spin systems to QFTs in periodic geometries, we extend the correspondence to systems with boundaries, where boundary conditions play a crucial role in shaping the dynamics. Focusing on Majorana fermions, we derive the allowed boundary conditions from the requirement of inner product conservation and formulate their realization in spin systems. The corresponding spin model is shown to reproduce boundary conditions of QFT accurately when a free function in the spin model is appropriately chosen. As an explicit demonstration, we analyze a flat spacetime example, comparing spectra, mode functions, and linear responses between the continuum and lattice descriptions. Our findings confirm that open spin systems can successfully replicate QFT dynamics with boundaries. 
}

\begin{document}
\maketitle

\section{Introduction}

Quantum field theories (QFTs) in curved spacetimes provide a versatile framework for studying physical systems ranging from quantum gravity to condensed matter physics \cite{BD,PhysRev.135.A1505,Muk,PhysRevLett.106.076601,PhysRevResearch.1.032006,CHERNODUB20221,10.1093/ptep/ptu162}.
The presence of boundaries in such systems introduces rich physical phenomena, including edge modes and modified dynamics, which are crucial for understanding topological properties and quantum phase transitions.
From the perspective of QFT itself, boundaries are known to give rise to a variety of intriguing effects, such as particle production induced by moving mirrors \cite{Davies:1976hi, Davies:1977yv} and the (dynamical) Casimir effect \cite{Moore:1970tmc,Yablonovitch:1989zza,Schwinger1992CasimirDielectrics, Dodonov:2010zza}.
More recently, boundary conformal field theories (BCFTs) have attracted significant attention, together with their holographic dual descriptions \cite{Cardy:1984bb,Cardy:1989ir,Takayanagi:2011zk}.
Spin systems, owing to their experimental feasibility and theoretical tractability, offer a powerful platform for simulating such boundary QFTs, enabling the study of complex field-theoretic phenomena in a controlled setting.
By taking the continuum limit of spin systems, one can map their dynamics to those of QFTs, thereby providing new insights into both theoretical and experimental contexts.

In our previous work \cite{Kinoshita:2024ahu}, we demonstrated that spin systems can serve as quantum simulators for QFTs in arbitrary two-dimensional curved spacetimes with periodic spatial coordinates. 
In a subsequent study, we applied the formalism to investigate the thermal properties of an inflationary universe \cite{kinoshita2025spinsystemsquantumfield}. 
However, many physical systems such as those in condensed matter or quantum gravity feature boundaries that significantly alter their behavior. 
These boundaries necessitate specific conditions that must be carefully incorporated into the spin system to ensure consistency with the corresponding QFT. 
This paper extends our prior framework to spin systems with boundaries, focusing on the implementation and implications of boundary conditions in the continuum limit.

We explore an open spin system with boundaries can accurately reproduce the dynamics of QFTs in two-dimensional spacetimes with boundaries. 
We derive the appropriate boundary conditions for the continuum theory and map them to the discrete lattice of the spin system, ensuring that the continuum limit preserves the essential physical properties. 
Using a flat spacetime example, we compare the continuum and discrete theories by analyzing their spectra and mode structures to validate our framework. 
Our results show that the proper choice of parameters in the spin system correctly replicates the QFT boundary conditions.

The paper is organized as follows. 
In Section \ref{sec:Review}, we review the previous work on the mapping from the spin systems to the quantum field theories in curved spacetimes. 
Section \ref{sec:BCforMajorana} explores possible boundary conditions for Majorana fermions based on the preservation of the inner product. 
Section \ref{sec:BCforSpinSystem} derives the proper boundary condition for the spin system which reproduces the boundary condition obtained in Sec.~\ref{sec:BCforMajorana} in the continuum limit. 
Section \ref{sec:FlatSpaceTime} shows the results for the flat space time. 
Section \ref{sec:Comparison} compares the continuum and discrete theories, including spectra and mode functions and linear response.  
In this Section \ref{sec:nonuniform_p}, 
we investigate the case where the free function in the spin Hamiltonian is spatially non-uniform. 
While its boundary values are fixed by the continuum boundary conditions, its bulk profile remains unconstrained; we analyze how such non-uniform choices affect the spectrum and discuss the emergence of local doubler excitations. 
Section \ref{sec:summary} is devoted to the summary and the discussion.

\section{Spin systems quantum simulators of quantum field theories in curved spacetimes}\label{sec:Review}

In Ref.~\cite{Kinoshita:2024ahu}, we demonstrated that the quantum field theory (QFT) of Majorana fermions in arbitrary two-dimensional curved spacetimes can be obtained by taking the continuum limit of a spin system with proper choice of parameters. 
In this section, we briefly review that study. We focus here on the case where the spatial coordinate is periodic. For a finite interval, one must take into account boundary conditions, which will be discussed in the following sections. 

We adopt the $(-+)$ signature for the two-dimensional Lorentzian metric $g_{\mu\nu}$. The two-dimensional gamma matrices in the Majorana representation are given by
\begin{equation}
    \Cliff^0 = i\sigma^y =
    \begin{pmatrix} 
        0 & 1 \\
        -1 & 0
    \end{pmatrix}, \quad 
    \Cliff^1 = \sigma^z =
    \begin{pmatrix} 
        1 & 0 \\
        0 & -1
    \end{pmatrix}.
\end{equation}
These satisfy the Clifford algebra $\{\Cliff^i, \Cliff^j\} = 2\eta^{ij} E$ with $\eta^{ij}= \mathrm{diag}(-1,1)$, where $E$ is the $2\times 2$ identity matrix. 
We introduce zweibeins $e^i_\mu$ satisfying
\begin{equation}
    g^{\mu\nu} e^i_\mu e^j_\nu = \eta^{ij}.
\end{equation}
We use Latin indices $i, j, k, \ldots$ for local Lorentz coordinates, which run over $0$ and $1$, and Greek indices $\mu, \nu, \rho, \ldots$ for general spacetime coordinates.
The gamma matrices with spacetime indices are defined as $\Cliff^\mu = e^\mu_i \Cliff^i$.

The action for a Majorana fermion with mass $m$ in two-dimensional curved spacetime is given by~\cite{Polyakov:1981re,deLacroix:2023uem,Erbin,Kinoshita:2024ahu}\footnote{
Compared to our previous paper~\cite{Kinoshita:2024ahu}, we have rescaled the Majorana field as $\psi \to \psi/\sqrt{2}$.
}
\begin{equation}
    S = -\frac{i}{2} \int d^2x\, \sqrt{-g}\, \bar{\psi} (\slashed{\nabla} - m) \psi \ ,\label{eq:Majorana_action}
\end{equation}
where $\bar{\psi} = \psi^\dagger \Cliff^0$, and $\psi = (a, b)^T$ is a two-component spinor field composed of real Grassmann-valued variables, satisfying $a^\dagger = a$ and $b^\dagger = b$. The covariant derivative for spinor fields in the zweibein formalism is defined as
\begin{equation}
    \nabla_\mu \psi = \left( \partial_\mu + \frac{1}{4} \omega_{\mu ij} \Cliff^{[i} \Cliff^{j]} \right) \psi ,
    \label{eq:covariantD_spinor}
\end{equation}
where $\omega_{\mu ij}$ is the spin connection. The explicit expression for the spin connection is not required in this paper since the spin connection vanishes in the action~\cite{deLacroix:2023uem,Erbin,Kinoshita:2024ahu}. 
The corresponding equation of motion is
\begin{equation}
    (\slashed{\nabla} - m) \psi = 0\ . \label{EOM}
\end{equation}
Also, taking the Hermitian conjugate of the equation of motion yields
\begin{equation}
    \nabla_i \bar{\psi} \Cliff^i + m \bar{\psi} = 0\ , \label{EOM2}
\end{equation}
where we have used the identity $(\Cliff^i)^\dagger = \Cliff^0 \Cliff^i \Cliff^0$.

In two dimensions, the most general metric is given by 
\begin{equation}
    ds^2=-\alpha(t,x)^2dt^2+\gamma(t,x)^2(dx-\beta(t,x)dt)^2 .
    \label{generalmetric}
\end{equation}
We assume that the spatial coordinate is periodic, $x \sim x + \ell$, and that the Majorana field satisfies the anti-periodic boundary condition, $\psi(x + \ell) = -\psi(x)$.
We introduce zweibeins and their dual vectors as
\begin{equation}
\begin{split}
 &e^0_\mu dx^\mu =\alpha dt\ ,\quad 
    e^1_\mu dx^\mu =\gamma (dx -\beta dt)\ ,\\
&e_0^\mu \partial_\mu =\frac{1}{\alpha}\partial_t + \frac{\beta}{\alpha}\partial_x\ ,\quad 
    e_1^\mu \partial_\mu =\frac{1}{\gamma}\partial_x \ .
\end{split}
\label{zwei0}
\end{equation}

We define the complex variable $\Psi$ as
\begin{equation}
\Psi = \frac{1}{\sqrt{2}}\gamma^{1/2} e^{-i\zeta /2} (b - ia)\ ,
\label{Psidef}
\end{equation}
where $\zeta$ is an arbitrary real function appearing in the phase of $\Psi$.
Although this phase can be eliminated by a field redefinition, we retain it here for convenience in later discussions.
From the action~(\ref{eq:Majorana_action}), we obtain the Hamiltonian density for a Majorana fermion in general two-dimensional metric as 
\begin{equation}
\begin{split}
\mathcal{H}=&-\frac{\alpha}{2 \gamma}\cos\zeta (\Psi^\dagger \partial_x \Psi^\dagger-\Psi \partial_x \Psi) 
+ i\frac{\alpha}{2 \gamma}\sin\zeta (\Psi^\dagger \partial_x \Psi^\dagger + \Psi \partial_x \Psi)
\\
&  -\frac{i\beta}{2} (\Psi^\dagger  \partial_x \Psi +\Psi \partial_x \Psi^\dagger)
-\left[m \alpha - \frac{1}{2}(\partial_t \zeta + \beta\partial_x\zeta) \right]\Psi^\dagger \Psi \ .
\end{split}
\label{MajH}
\end{equation}

The above Hamiltonian is obtained by taking the continuum limit of a spin model as follows.
We consider a Hamiltonian of the spin model given by
\begin{equation}
\begin{split}
    H=-\frac{1}{4\varepsilon} \sum_{j=1}^L\bigg\{&
    \left(\frac{\alpha(t,x_j)}{\gamma(t,x_j)} \cos\zeta(t,x_j) + p(t,x_j)\right)\sigma^x_j \sigma^x_{j+1}\\
    - &\left(\frac{\alpha(t,x_j)}{\gamma(t,x_j)} \cos\zeta(t,x_j) - p(t,x_j)\right)\sigma^y_j \sigma^y_{j+1} \\
    - &\left(\beta(t,x_j) + \frac{\alpha(t,x_j)}{\gamma(t,x_j)} \sin\zeta(t,x_j)\right) \sigma^x_j \sigma^y_{j+1} \\
    + &\left(\beta(t,x_j) - \frac{\alpha(t,x_j)}{\gamma(t,x_j)} \sin\zeta(t,x_j)\right) \sigma^y_j \sigma^x_{j+1}\\
    + &\big[2p(t,x_j) - \varepsilon \big(\partial_x p(t,x_j) + 2m\alpha(t,x_j)\\
    &\hspace{0.3cm} - \partial_t\zeta(t,x_j) - \beta(t,x_j)\partial_x\zeta(t,x_j)\big)\big] \sigma^z_j \bigg\} ,
    \label{IsingforQFT}
\end{split}
\end{equation}
where $L$ is the number of sites ($\sigma^{x,y,z}_{L+1}=\sigma^{x,y,z}_1$) and $\varepsilon$ is the lattice spacing, defined by $\varepsilon = \ell / L$.
We assign spatial coordinates to the spin sites as 
\begin{equation}
    x_j = \varepsilon (j-j_0)\ ,
    \label{xjdef}
\end{equation}
where $j_0$ is an arbitrary real constant that fixes the choice of coordinate origin (i.e., which site is taken to be at $x=0$).%
\footnote{
We currently assume that sites are located at uniformly spaced coordinates with interval $\varepsilon$. This assumption is made without loss of generality, because any non-uniform spacing can be absorbed into the metric function $\gamma(t,x)$ and $\beta(t,x)$ via a suitable coordinate transformation.
}
This dictionary allows us to construct a corresponding spin system for a quantum field theory in a curved spacetime, once the metric components $\alpha$, $\beta$, and $\gamma$ are specified. 
Here, $p(t,x)$ is an arbitrary real function.
After the Jordan-Wigner transformation:
\begin{equation}
    \frac{\sigma^x_j+i\sigma^y_j}{2}=\prod_{l=1}^{j-1}(1-2c_l^\dagger c_l) c_j\ ,\quad 
    \sigma_j^z=1-2c_j^\dagger c_j\ ,
    \label{JW}
\end{equation}
the Hamiltonian of the spin system is rewritten as
\begin{equation}
\begin{split}
 H=&- \frac{1}{2\varepsilon} \sum_{j=1}^{L}\bigg\{ \frac{\alpha(t,x_j)}{\gamma(t,x_j)}\cos\zeta(t,x_j)
 ( c_{j+1}c_j +c_j^\dagger c_{j+1}^\dagger)\\
& + i\frac{\alpha(t,x_j)}{\gamma(t,x_j)}\sin\zeta(t,x_j)
 ( c_{j+1}c_j - c_j^\dagger c_{j+1}^\dagger)\\
& + p(t,x_j)( c_j^\dagger c_{j+1}+ c_{j+1}^\dagger c_j)
 \\
 &+ i\beta(t,x_j)( c_j^\dagger c_{j+1}- c_{j+1}^\dagger c_j)\\
&+\frac{1}{2}\big[2p(t,x_j) - \varepsilon\big(\partial_x p(t,x_j) + 2m\alpha(t,x_j)\\
&\hspace{0.2cm} - \partial_t\zeta(t,x_j) - \beta(t,x_j)\partial_x\zeta(t,x_j)\big)\big]
(1-2c_j^\dagger c_j)\bigg\} ,
 \label{IsingforQFTc}
\end{split}
\end{equation}
where the fermionic operators satisfy $\{c_i,c_j^\dagger\}=\delta_{i,j}$ and $\{c_i,c_j\}=0$. 
The spin system is mapped into the fermion system. We identify the fermionic operator $c_j$ and Majorana field $\Psi(x_j)$ as
\begin{equation}
    \Psi(x_j) = \frac{c_j}{\sqrt{\varepsilon}}\ .
    \label{Psi_and_c}
\end{equation}
Under this identification, taking the limit $L \to \infty$ with $\ell/L\to 0$ yields the Hamiltonian of the quantum field theory~(\ref{MajH}). In the present case of periodic spatial coordinates, the dependence on the real function $p(t,x)$ cancels out in the continuum limit, so that the resulting quantum field theory is independent of its choice.

\section{Allowed boundary conditions for Majorana fermions and their realization in spin systems}
\subsection{Majorana fermions with open boundaries}\label{sec:BCforMajorana}

From this point on, we consider the quantum field theory of Majorana fermions in a spacetime with boundaries. In this case, it is necessary to examine the boundary conditions for the Majorana fermions. We determine the allowed boundary conditions based on the requirement that the inner product be preserved.

Let us assume that both $\psi_1$ and $\psi_2$ satisfy the equation of motion~\eqref{EOM}.  
On a spacelike hypersurface $\Sigma$, we define the inner product for Majorana fields as  
\begin{equation}
    (\psi_1, \psi_2) = \int_\Sigma d\Sigma\, n_\mu \bar{\psi}_1 \Cliff^\mu \psi_2 , \label{inprod}
\end{equation}  
where $d\Sigma$ is the surface element of $\Sigma$, and $n^\mu$ is the future-directed unit normal vector to $\Sigma$.  
From Eqs.~\eqref{EOM} and \eqref{EOM2}, it follows that
\begin{equation}
    \nabla_\mu \left( \bar{\psi}_1 \Cliff^\mu \psi_2 \right) = 0\ . \label{eq:conservation_prod}
\end{equation}
This implies that the inner product~\eqref{inprod} is independent of the choice of spacelike hypersurface $\Sigma$ if the spacetime has no spatial boundaries.

Next, we consider a spacetime region $V$ enclosed by the constant-time surfaces $t = t_1$ and $t = t_2$, together with a time-like boundary $S$. 
Integrating Eq.~(\ref{eq:conservation_prod}) over $V$, we obtain 
\begin{equation}
    (\psi_1, \psi_2)|_{t=t_2} - (\psi_1, \psi_2)|_{t=t_1}
    = - \int_{S} dS\, m_\mu \bar{\psi}_1 \Cliff^\mu \psi_2 \ ,
\end{equation}
where $dS$ denotes the surface element on $S$, and $m_\mu$ is the outward-pointing unit normal vector to $S$.
If we take the spatial boundaries at 
constant-$x$ surfaces ($x=0$ and $x=\ell$), the surface element is given by $dS = \sqrt{\alpha^2 - \beta^2 \gamma^2}\, dt$, and the unit normal vector to the boundary is $\pm m_\mu=\delta^x_\mu/\sqrt{g^{xx}}=\alpha\gamma \delta^x_\mu/\sqrt{\alpha^2-\beta^2\gamma^2}$.
Using the expression $\Cliff^x = e^x_i\Cliff^i = (\beta/\alpha) \Cliff^0 + (1/\gamma) \Cliff^1$, the boundary contribution becomes
\begin{multline}
    (\psi_1, \psi_2)\big|_{t = t_2} - (\psi_1, \psi_2)\big|_{t = t_1}
    = \int dt\, \bar{\psi}_1 \left( \beta \gamma\, \Cliff^0 + \alpha\, \Cliff^1 \right) \psi_2 \big|_{x = \ell}^{x = 0} \\
    = -\int dt\, \psi_1^\dagger \left( \beta \gamma + \alpha\, \sigma^x \right) \psi_2 \big|_{x = 0}^{x = \ell} .
    \label{bdryterm}
\end{multline}



Here, we identify boundary conditions that ensure the conservation of the inner product. The following discussion is based on Ref.~\cite{Witten:2015aoa}. (See also \cite{Hashimoto:2016kxm, Al-Hashimi:2016vfd}.) 
At the boundaries, we assume that the Majorana field satisfies
\begin{equation}
    \psi = M\psi\ , \label{psicond}
\end{equation}  
where $M$ is a $2\times 2$ real matrix. Defining $A = \beta \gamma + \alpha \sigma^x$, we can rewrite the integrand of Eq.~\eqref{bdryterm} as  
\begin{equation}
\begin{split}
    2\psi_1^\dagger A \psi_2 &= \psi_1^\dagger A \psi_2 + \psi_1^\dagger A \psi_2 \\
    &= \psi_1^\dagger A M \psi_2 + \psi_1^\dagger M^T A \psi_2\\
    &= \psi_1^\dagger(AM+M^T A) \psi_2\ .
\end{split}
\end{equation}  
Therefore, if the condition
\begin{equation}
    AM+M^T A = 0\ 
    \label{Mcond}
\end{equation}
is satisfied at the boundaries, 
then the right-hand side of Eq.~(\ref{bdryterm}) vanishes, and the inner product is conserved.
To determine the explicit expression of the matrix $M$, we expand it in terms of the Pauli matrices as  
\begin{equation}
    M = c_0 + c_x \sigma^x + c_y i \sigma^y + c_z \sigma^z\ ,\quad (c_0, c_x, c_y, c_z \in \mathbb{R})\ . 
\end{equation}  
Substituting this expression into Eq.~\eqref{Mcond}, we obtain  
\begin{equation}
    \frac{c_y}{c_z}=\frac{\beta \gamma}{\alpha}\ ,\quad c_0=c_x=0 .
\end{equation}  
Thus, the real matrix $M$ is given by  
\begin{equation}
    M=\lambda (\beta\gamma i\sigma^y + \alpha \sigma^z)
    =
    \lambda 
    \begin{pmatrix}
        \alpha & \beta \gamma \\
        -\beta \gamma & -\alpha
    \end{pmatrix}\ ,
    \label{Mform}
\end{equation}  
where $\lambda$ is a real constant.
The equation of motion for the Majorana field is a first-order differential equation. Therefore, if all components of the field vanish at the boundary, then the field must vanish throughout the entire spacetime, rendering the dynamics trivial.
From Eq.~(\ref{psicond}), the condition for the field to have nontrivial values at the boundary is $\det(1 - M) = 0$. This condition determines the constant $\lambda$ as
\begin{equation}
   \lambda = \frac{s}{\sqrt{\alpha^2-\beta^2\gamma^2}}\ ,
   \label{lambda}
\end{equation}  
where $s=1$ or $-1$. 
According to this argument, only two types of boundary conditions are allowed.
Note that the $s$ can independently take the values $\pm 1$ at each end of the domain. 
Since the boundary is located at $x = \mathrm{constant}$, its tangent vector is $\partial_t$. Therefore, the condition for the boundary to be time-like is $g_{\mu\nu}(\partial_t)^\mu(\partial_t)^\nu=g_{tt} = -\alpha^2 + \beta^2 \gamma^2 < 0$. This implies that the parameter $\lambda$ introduced above is always real.

In terms of $\Psi$ defined in Eq.~(\ref{Psidef}), the boundary condition for the Majorana field is given by
\begin{equation}
     \Psi=- \frac{s}{\sqrt{\alpha^2-\beta^2\gamma^2}} 
     (\alpha e^{-i\zeta}\Psi^\dagger + i \beta \gamma \Psi)\  \qquad (s=\pm 1) .
     \label{Psibc}
\end{equation}
From the above equation, it follows that $\Psi^\dagger \Psi=0$ at boundaries.%
\footnote{In the chiral representation (see appendix~\ref{app:lattice_fermion}), these condtions are rewritten as 
$\sqrt{\alpha + \beta\gamma}\chi^+ = - s \sqrt{\alpha -\beta\gamma}\chi^-$, where 
$\pmat{\chi^+ \\ \chi^-} \equiv \dfrac{1}{\sqrt{2}}\pmat{1 & 1 \\ -1 & 1} \psi$. }

For example, in the case of flat spacetime, i.e., when $\alpha = \gamma = 1$ and $\beta = 0$, the matrix becomes
\begin{equation}
    M=s\pmat{1 & 0 \\ 0 & -1}\ .
\end{equation}
If we write the Majorana field as a two-component spinor $\psi = (a, b)^T$, then the boundary conditions are written as 
\begin{equation}
    b=0\quad (\textrm{for }s=1)\ ,\qquad a=0 \quad (\textrm{for }s=-1)\ .
\end{equation}

\subsection{Corresponding boundary conditions in spin systems}\label{sec:BCforSpinSystem}

Let us consider the spin system corresponding to QFT of Majorana fermions in a space with allowed boundary conditions. In the bulk, the procedure for taking the continuum limit of the spin system is the same as in the case with periodic boundary conditions. Therefore, the corresponding spin system is essentially given by Eq.~(\ref{IsingforQFT}) or Eq.~(\ref{IsingforQFTc}). Here, our goal is to identify the conditions under which the continuum limit of the spin system correctly reproduces the boundary conditions of QFT.

From the Jordan-Wigner fermion representation given in Eq.~(\ref{IsingforQFTc}), the Heisenberg equations for $c_j$ is
\begin{multline}
    i\dot{c}_j
    =-\frac{1}{2\varepsilon}\bigg\{\frac{\alpha_j}{\gamma_j}e^{-i\zeta_j} c_{j+1}^\dagger-\frac{\alpha_{j-1}}{\gamma_{j-1}}e^{-i\zeta_{j-1}} c_{j-1}^\dagger\\
    +p_j c_{j+1}+p_{j-1}c_{j-1}
    +i\beta_j c_{j+1}-i\beta_{j-1}c_{j-1}\\
    -[2p_j - \varepsilon(\partial_x p_j + 2m\alpha_j- \partial_t\zeta_j - \beta_j\partial_x\zeta_j)]c_j\bigg\}
\label{eq:Heisenberg_ci}
\end{multline}
Here, we use the shorthand notation $\alpha(t,x_j)=\alpha_j$ and $\beta(t,x_j)=\beta_j$, and so on.
Within the range $2 \leq j \leq L - 1$, the above expression is valid without any additional requirement. To include the cases $j = 1$ and $j = L$ in a unified manner, one can formally introduce virtual sites at $j = 0$ and $j = L+1$. 
Since these sites are purely auxiliary, suitable conditions must be imposed on the associated operators to ensure that they do not affect the dynamics governed by Eq.~(\ref{eq:Heisenberg_ci}), namely,
\begin{equation}
\begin{split}
    &c_{0}=\frac{1}{p_0\gamma_0}(\alpha_{0}e^{-i\zeta_{0}} c_{0}^\dagger+i\beta_{0}\gamma_0c_{0})\ ,\\
    &c_{L+1}=-\frac{1}{p_L\gamma_L}(\alpha_Le^{-i\zeta_L} c_{L+1}^\dagger+i\beta_L\gamma_L c_{L+1})\ .
\end{split}
\end{equation}
According to Eq.~(\ref{Psi_and_c}), in the continuum limit, the above conditions take the following form:
\begin{equation}
    \Psi|_{x=0}=\frac{1}{p \gamma }(\alpha e^{-i\zeta} \Psi^\dagger+i\beta \gamma \Psi)\bigg|_{x=0}\ ,\quad
    \Psi|_{x=\ell}=-\frac{1}{p \gamma }(\alpha e^{-i\zeta} \Psi^\dagger+i\beta \gamma \Psi)\bigg|_{x=\ell} .
\end{equation}
By comparing these expressions with Eq.~(\ref{Psibc}), 
we find that, in order to reproduce the allowed boundary conditions in QFT based on the present spin system, the boundary values of $p$ must satisfy the following conditions: 
\begin{equation}
    p|_{x=0}=-s \sqrt{\frac{\alpha^2}{\gamma^2}-\beta^2}\,\Bigg|_{x=0}\ ,\quad 
    p|_{x=\ell}= s' \sqrt{\frac{\alpha^2}{\gamma^2}-\beta^2}\,\Bigg|_{x=\ell}\ ,
    \label{pbc}
\end{equation}
where $s = \pm 1$ and $s' = \pm 1$ correspond to the signs that appear in the boundary conditions~(\ref{Psibc}) at the two ends in QFT. It is possible to assign distinct signs to each boundary.
When considering a quantum field theory with boundaries, constraints are imposed only on the boundary values of the function $p(t,x)$. 
This implies that the function $p(t,x)$ can be freely chosen in the bulk, provided that it is sufficiently smooth near the boundaries and its boundary values satisfy the condition stated above.
We will demonstrate this explicitly in Sec.~\ref{sec:nonuniform_p}.



\section{Flat spacetime example}\label{sec:FlatSpaceTime}

\subsection{Continuum theory}
\label{Conttheory}
We consider Majorana fermions in $0\leq x \leq \ell$. The metric is flat: $ds^2=-dt^2+dx^2$. 
The equation of motion for the Majorana field is $(\slashed{\partial}-m)\psi=0$. Writing the Majorana field as $\psi=(a,b)^T$, we obtain the Hamiltonian as
\begin{equation}
    H=-i\int^\ell_0 dx ( ab'+m ab )\ .
\end{equation}
The equations of motion are given by
\begin{equation}
    \dot{a}=-b'-mb\ ,\quad \dot{b}=-a'+ma\ ,
    \label{EOMab}
\end{equation}
where ${}^\cdot=\partial_t$ and ${}'=\partial_x$.
As mentioned in Sec.~\ref{sec:BCforMajorana}, two kinds of boundary conditions, distinguished by their signs, can be independently allowed at each boundary.
Here, we impose different types of boundary conditions at the boundaries $x=0$ and $x=\ell$, respectively, as
\begin{equation}
    a|_{x=0}=0\ ,\quad b|_{x=\ell}=0\ .
    \label{bc1}
\end{equation}
These boundary conditions correspond to setting $s = -1$ at $x = 0$ and $s = 1$ at $x = \ell$ in Eqs.~(\ref{Mform}) and (\ref{lambda}). 
This particular choice corresponds to the case of the spin system with $p=1$, as will be shown later.
In appendix~\ref{Majoranaflat}, the equations of motion (\ref{EOMab}) are solved under the boundary conditions~(\ref{bc1}). As a result, the solution is given as 
\begin{equation}
\begin{split}
&a(t,x)=i \sum_{n=1}^\infty f_n (x) (e^{-i\omega_n t}\gamma_n - e^{i\omega_n t}\gamma_n^\dagger)\ ,\\
&b(t,x)=\sum_{n=1}^\infty g_n(x) (e^{-i\omega_n t}\gamma_n + e^{i\omega_n t}\gamma_n^\dagger)\ ,
\end{split}
\label{absol}
\end{equation}
where the mode functions $f_n(x)$ and $g_n(x)$ satisfy $f_n(0)=g_n(\ell)=0$.
For $n\geq 2$ (excluding the edge mode at $n = 1$, to be introduced shortly), the eigenfrequency is given by $\omega_n=\sqrt{k_n^2+m^2}$, where   
the wave number $k_n$ is given by the positive solution of 
\begin{equation}
    \tan k_n \ell = \frac{1}{m\ell} k_n \ell \ .
\end{equation}
This equation admits infinitely many solutions, and we label the positive ones in ascending order as $k = k_n$ for $n = 2, 3, \dots$  
(the solution asymptotes to $k_n\sim \frac{\pi}{2}+\pi (n-1)$ for large $k$). 
Corresponding mode functions are given by
\begin{equation}
\begin{split}
    &f_n(x)=\left(\ell - \frac{m}{\omega_n^2}\right)^{-1/2}\sin k_n x\ ,\\
    &g_n(x)=\left(\ell - \frac{m}{\omega_n^2}\right)^{-1/2}
    \frac{1}{\omega_n}(k_n\cos k_n x-m \sin k_n x)
    \ .
\end{split}
\label{modefns}
\end{equation}
Depending on the values of the mass $m$ and the domain length $\ell$, there may exist, in addition to the above modes, a mode localized at the boundary. We refer to this as the \emph{edge mode}, and assign $n = 1$ to it. The eigenfrequency of the edge mode is given by $\omega_1=\sqrt{m^2-\kappa^2}$, where $\kappa$ is the positive solution of 
\begin{equation}
    \tanh \kappa\ell = \frac{1}{m\ell} \kappa \ell\ .
\end{equation}
This equation has a non-trivial solution within $0<\kappa<m$ only if $m\ell > 1$, and $\kappa$ approaches $m$ in the limit $m\ell \to \infty$.
The mode functions for the edge mode are  
\begin{equation}
\begin{split}
    &f_1(x)=\left\{\frac{1}{m}\left(1-\frac{\ell \omega_1^2}{m}\right)\right\}^{-1/2}\frac{\omega_1}{m} \sinh\kappa x\ ,\\
    &g_1(x)=\left\{\frac{1}{m}\left(1-\frac{\ell \omega_1^2}{m}\right)\right\}^{-1/2}\frac{1}{m}(\kappa \cosh \kappa x-m\sinh \kappa x)\ .
\end{split}
\label{edgemodefn}
\end{equation}
The edge mode of the Majorana field has been extensively studied in the literature, such as in Ref.~\cite{Kitaev01} and we will not go into detail here. Instead, we will confirm that the spin system provides a good approximation to the field theory only when the arbitrary function $p$ is chosen as in Eq.~(\ref{pbc}), for both the edge and bulk modes.
For $m\ell < 1$, there is no edge mode and the summation in Eq.~(\ref{absol}) starts from $n=2$. 
When $m\ell = 1$, there exists a linear mode whose mode function takes the form of a linear function. Since this case is not discussed in the main text, we omit it here. For details, see appendix~\ref{Majoranaflat}.

We can directly confirm the orthonormal relations for the mode functions introduced above as 
\begin{equation}
\begin{split}
&\int^\ell_0 dx f_n f_{n'} =  \int^\ell_0 dx g_n g_{n'}=\frac{1}{2}\delta_{nn'}\ .
\end{split}
\end{equation}
By using the above relations, we can also show
\begin{equation}
    \{\gamma_n,\gamma_{n'}^\dagger\}=\delta_{nn'}\ ,\quad \{\gamma_n,\gamma_{n'}\}=\{\gamma_n^\dagger,\gamma_{n'}^\dagger\}=0\ .
\end{equation}
The Hamiltonian is given by
\begin{equation}
     H=\sum_{n=1}^\infty
     \omega_n \gamma_n^\dagger \gamma_n + E_0\ ,
\end{equation}
where $E_0=-\sum_{n=1}^\infty \omega_n/2$. 
The ground state $|\Omega\rangle$ is defined by $\gamma_n |\Omega\rangle =0$ ($n=1,2,\dots$).


\subsection{Discrete theory}
\label{Dtheory}
In the previous subsection, we have discussed the QFT of a Majorana fermion in a flat space with a finite domain. As the spin-system counterpart of this setup, we now introduce the following Hamiltonian:
\begin{equation}
    H=-\frac{1}{4\varepsilon} \bigg\{\sum_{j=1}^{L-1}\bigg[
    (1 + p)\sigma^x_j \sigma^x_{j+1}-(1-p)\sigma^y_j \sigma^y_{j+1} \bigg]
    + 2(p - m\varepsilon) \sum_{j=1}^{L} \sigma^z_j \bigg\} .
\end{equation}
It can be obtained by substituting the metric functions of flat spacetime, $\alpha(t,x)=\gamma(t,x)=1$ and $\beta(t,x)=0$, into Eq.~(\ref{IsingforQFT}).
Although Eq.~(\ref{IsingforQFT}) contains an arbitrary function $\zeta(t,x)$ and a function $p(t,x)$ that is arbitrary in the bulk but constrained at the boundaries, we set $\zeta(t,x)=0$ and assume $p(t,x)$ to be constant to simplify subsequent analysis.
Substituting $\alpha(t,x) = \gamma(t,x) = 1$ and $\beta(t,x) = 0$ into Eq.~(\ref{pbc}), we obtain $p(t,x)|_{x=0} = p(t,x)|_{x=\ell} = 1$. In other words, if $p$ is taken to be constant, only $p = 1$ yields the ``correct'' boundary condition that reproduces the target QFT in the continuum limit. From the perspective of the spin system itself, however, arbitrary values of $p$ still appear to be admissible. Here, in order to investigate how boundary conditions with $p\neq 1$ influence the behavior of the spin system, we will leave $p$ as an arbitrary constant in the following calculations.
By the Jordan-Wigner transformation, the above Hamiltonian becomes
\begin{equation}
\label{HamiltonianJW}
\begin{split}
 H=&- \frac{1}{2\varepsilon}\bigg\{ \sum_{j=1}^{L-1}\bigg[
 c_{j+1}c_j +c_j^\dagger c_{j+1}^\dagger 
 + p( c_j^\dagger c_{j+1}+ c_{j+1}^\dagger c_j)\bigg]
+(p - m \varepsilon) \sum_{j=1}^{L} (1-2c_j^\dagger c_j) \bigg\}\\
=&\frac{1}{2}\pmat{\bm{c}^{T} & \bar{\bm{c}}^{T}}
\pmat{A & -S \\ S & -A} \pmat{\bm{c} \\ \bar{\bm{c}}}\ ,
\end{split}
\end{equation}
where $\bm{c}=(c_1,\dots,c_L)^T$ and $\bar{\bm{c}}=(c_1^\dagger,\dots,c_L^\dagger)^T$. We have introduced the symmetric matrix $S$ and anti-symmetric matrix $A$ as
\begin{equation}
    S=-\frac{1}{2\varepsilon}\left[2(m\varepsilon-p)E + p(E_++E_-)\right]\ ,\quad 
    A=\frac{1}{2\varepsilon}(E_+-E_-)\ ,
\end{equation}
where $E$ is the identity matrix and $E_\pm$ are upper and lower shift matrices defined by
\begin{equation}
E_+=
    \begin{pmatrix}
0\, & 1 &   &        &        \\
  & 0\, & 1 &        &        \\
  &   & \ddots & \ddots &        \\
  &   &        & 0\, & 1 \\
  &   &        &   & 0\,
\end{pmatrix} \ ,\quad 
    E_-=E_+^T\ .
    \label{Epmdef}
\end{equation}

Introducing Hermitian operators $\bm{a}$ and $\bm{b}$ through $\bm{c}=(\bm{b}-i\bm{a})/\sqrt{2}$ and $\bar{\bm{c}}=(\bm{b}+i\bm{a})/\sqrt{2}$, the Hamiltonian is rewritten as
\begin{equation}
    H=\frac{1}{2}\pmat{\bm{a}^T & \bm{b}^T}
M \pmat{\bm{a} \\ \bm{b}}\ ,\quad M=i\pmat{0 & B^T\\
    -B & 0}\ ,
    \label{Hab}
\end{equation}
where $B=S+A$ and $B^T = S-A$.
The corresponding Heisenberg equation of motion is given by
\begin{equation}
    \frac{d}{dt}\pmat{\bm{a} \\ \bm{b}}=-i M
    \pmat{\bm{a} \\ \bm{b}}\ .
    \label{abeqd}
\end{equation}
Note that $a_i$ and $b_i$ satisfy the following anti-commutation relations: $\{a_i, a_j\}=\{b_i, b_j\} = \delta_{ij}$ and $\{a_i, b_j\}=0$.
By diagonalizing the matrix $M$, we can immediately obtain the solution to the above equation.
Eigenvalue equation for the matrix $M$ is
\begin{equation}
    |\omega^ 2E - BB^T|=0\ . 
\end{equation}
Therefore, the eigenvalues always appear in pairs of positive and negative values. We denote them as
\begin{equation}
    \omega_1, \omega_2, \dots, \omega_L, -\omega_1,-\omega_2,\dots,-\omega_L\ ,
\end{equation}
with $\omega_j > 0$ ($j=1,2,\dots,L$). In the parameter regime considered in this paper, we have confirmed by explicit calculation that $\omega_j\neq 0$ for all $j$.
Let a $2L$-dimensional column vector $(i\bm{f}_j, \bm{g}_j)^T$  be an eigenvector of $M$ corresponding to the eigenvalue $\omega_j$. 
The $L$-dimensional column vectors $\bm{f}_j$ and $\bm{g}_j$ satisfy
\begin{equation}
    \omega_j^2 \bm{f}_j = B^TB\bm{f}_j\ ,\quad \bm{g}_j=\frac{1}{\omega_j}B \bm{f}_j\ .
    \label{fgeq}
\end{equation}
The procedure for obtaining the eigenvalues and eigenvectors of the matrix $M$ is as follows. First, according to the first equation in Eq.~(\ref{fgeq}), determine the eigenvalues $\omega_j^2$ and the corresponding eigenvectors $\bm{f}_j$ of $B^T B$. The values $\omega_j$ are then obtained by taking the square root of these eigenvalues. Next, by substituting $\bm{f}_j$ and $\omega_j$ into the second equation of Eq.~(\ref{fgeq}), one can compute $\bm{g}_j$. By stacking the two as $(i \bm{f}_j, \bm{g}_j)^T$,  one obtains the eigenvector of $M$. The eigenvector corresponding to the eigenvalue $-\omega_j$ is given by $(-i \bm{f}_j, \bm{g}_j)^T$. Since $B^T B$ is a real symmetric matrix, all components of $\bm{f}_j$ and $\bm{g}_j$ can be taken as real values.
We normalize the eigenvectors such that $\bm{f}^T \bm{f}=\bm{g}^T \bm{g}=1/2$. 
Defining the unitary matrix 
\begin{equation}
    U=\pmat{
    i\bm{f}_1 & \cdots &i\bm{f}_L & -i\bm{f}_1 & \cdots &-i\bm{f}_L\\
    \bm{g}_1 & \cdots &\bm{g}_L & \bm{g}_1 & \cdots &\bm{g}_L
    }\ ,
\end{equation}
we can diagonalize the matrix $M$ as
\begin{equation}
    U^\dagger M U = \textrm{diag}(\omega_1,\dots,\omega_L,-\omega_1,\dots,-\omega_L)\ .
\end{equation}
Therefore, the solution of Eq.~(\ref{abeqd}) is given by
\begin{equation}
    \pmat{\bm{a}\\\bm{b}}
    =U 
    \pmat{\gamma_1 e^{-i\omega_1 t}\\ \vdots \\ \gamma_L e^{-i\omega_L t}\\
    \gamma_1^\dagger e^{i\omega_1 t}\\ \vdots \\ \gamma_L^\dagger e^{i\omega_L t}}
    \label{absolvec}
\end{equation}
or equivalently, 
\begin{equation}
\begin{split}
    &a_j=i\sum_k f_{jk}( \gamma_k e^{-i\omega_k t} - \gamma_k^\dagger e^{i\omega_k t} )\ ,\\
    &b_j=\sum_k g_{jk}( \gamma_k e^{-i\omega_k t} + \gamma_k^\dagger e^{i\omega_k t} )\ ,
\end{split}
\end{equation}
where the components of vector quantities are written as $\bm{a}=(a_1,\dots,a_L)^T$, $\bm{b}=(b_1,\dots,b_L)^T$, $\bm{f}_j=(f_{1j},\dots,f_{Lj})^T$ and $\bm{g}_j=(g_{1j},\dots,g_{Lj})^T$. From the unitarity of the matrix $U$, we have
\begin{equation}
\sum_{k=1}^Lf_{ik}f_{jk}=\sum_{k=1}^Lf_{ki}f_{kj}=\sum_{k=1}^Lg_{ik}g_{jk}=\sum_{k=1}^Lg_{ki}g_{kj}=\frac{1}{2}\delta_{ij}\ .
\end{equation}
Using the above equations, we can confirm that 
\begin{equation}
    \{\gamma_j,\gamma_k^\dagger\}=\delta_{jk}\ ,\quad 
    \{\gamma_j,\gamma_k\}=\{\gamma_j^\dagger,\gamma_k^\dagger\}=0\ .
\end{equation}
Substituting Eq.~(\ref{absolvec}) in the Hamiltonian~(\ref{Hab}), we have
\begin{equation}
    H=\frac{1}{2}\sum_{j=1}^L \omega_j(\gamma_j^\dagger \gamma_j-\gamma_j\gamma_j^\dagger)=
    \sum_{j=1}^L \omega_j\gamma_j^\dagger \gamma_j + E_0\ ,
\end{equation}
where $E_0=-\sum_{j=1}^L \omega_j/2$ is the energy of the ground state. The ground state $|\Omega\rangle$ is defined by $\gamma_j|\Omega\rangle = 0$ for $j=1,2,\ldots,L$.

\section{Comparison of the continuum and discrete theories}\label{sec:Comparison}

\subsection{Spectra and mode functions}
\label{spec_and_modef}
The correspondence of quantum fields between the continuum and discrete theories is given by Eq.~(\ref{Psi_and_c}). In terms of real variables we have 
$a(x_j) = a_j/\sqrt{\varepsilon}$ and $b(x_j) = b_j/\sqrt{\varepsilon}$. 
Accordingly, the correspondence between the mode functions is
\begin{equation}
    f_{n}(x_j)=\frac{f_{jn}}{\sqrt{\varepsilon}}\ ,\quad 
    g_{n}(x_j)=\frac{g_{jn}}{\sqrt{\varepsilon}}\ .
\end{equation}

\begin{figure}
\begin{center}
\includegraphics[scale=0.5]{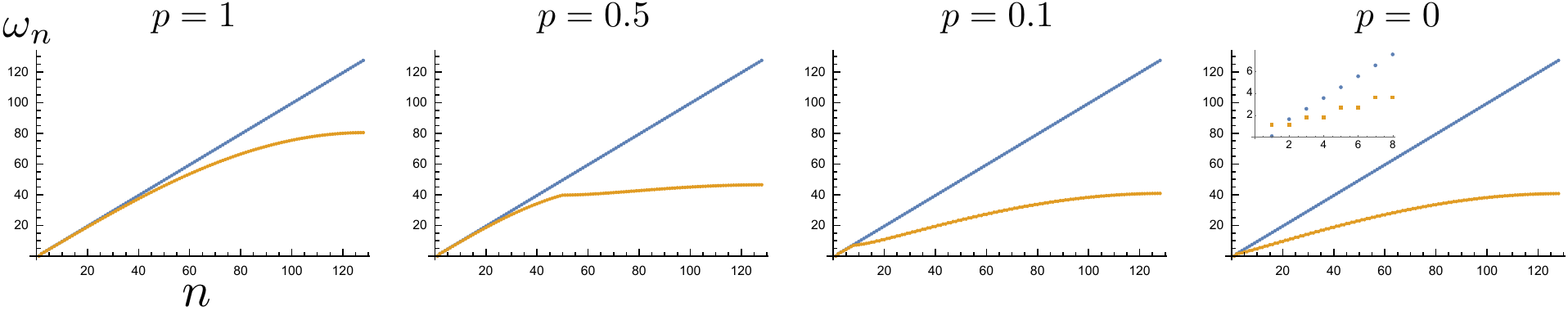}
\end{center}
\caption{Spectra of QFT (blue dots) and spin systems (orange dots) for $p = 1, 0.5, 0.1,$ and $0$. 
The parameters are set to $L = 128$, $\ell = \pi$, and $m = 1$.
The spectrum of the spin system typically begins to deviate from that of the QFT at the doubler mass scale given by $2p/\varepsilon - m$. }
\label{omega}
\end{figure}

Figure~\ref{omega} presents a comparison between the spectra $\omega_n$ of QFT and the corresponding spin system. We set the parameters as $L = 128$, $\ell = \pi$, and $m = 1$. Since $m\ell > 1$, in QFT, an edge mode exists as the first excited state. The panels, from left to right, correspond to $p = 1$, $0.5$, $0.1$, and $0$. (The spin system correctly reproduces the boundary conditions of QFT for $p=1$.) The blue dots represent the results from the QFT, while the orange dots represent those from the spin system. The wavelength of the $n$-th mode can be estimated as $\lambda_n \sim \ell / n$. For the spin system to serve as a good approximation to the continuum QFT, the wavelength must be much larger than the lattice spacing: $\lambda_n \gg \varepsilon = \ell / L$, i.e., $n\ll L$. 
In general, discrete fermion theories inevitably contain additional light modes whose wavelengths are of the order of the lattice spacing, known as doublers. In this case, the mass of the doubler can be estimated to be $2p/\varepsilon - m$. (See appendix~\ref{app:lattice_fermion}.) Therefore, the spectrum of the spin system is expected to deviate from that of the continuum QFT at this mass scale, due to the emergence of bulk excitations associated with the doubler. 
The panel for $p=1$ in Fig.~\ref{omega} confirms that the discrete theory reproduces the QFT spectrum well for $n\ll L$. As $p$ deviates from $1$, the discrepancy between the spectra of QFT and the spin system becomes more pronounced. In particular, for $p = 0$, degenerate pairs of eigenfrequencies appears.\footnote{Under periodic boundary conditions, the wave number $k$ becomes a good quantum number, and the dispersion relation in this case shows that the gap narrows at two points in momentum space, specifically at $k=0$ and $k=\pi$. In this sense, the theory at $p=0$ has doublers, and this degeneracy can be regarded as originating from them.} This suggests that the mode index $n$ for the spin system should perhaps be assigned as $n|_{\textrm{QFT}} \sim 2n|_{\textrm{Spin}}$. Even with such a reinterpretation of mode numbers, one finds that the eigenvalue corresponding to the edge mode of QFT is absent in the spin system when $p=0$. In the language of the Kitaev chain model, this can be understood as the system
lying outside the topological phase (see Sec.~\ref{kitaevchain}).

\begin{figure}
\begin{center}
\includegraphics[scale=0.95]{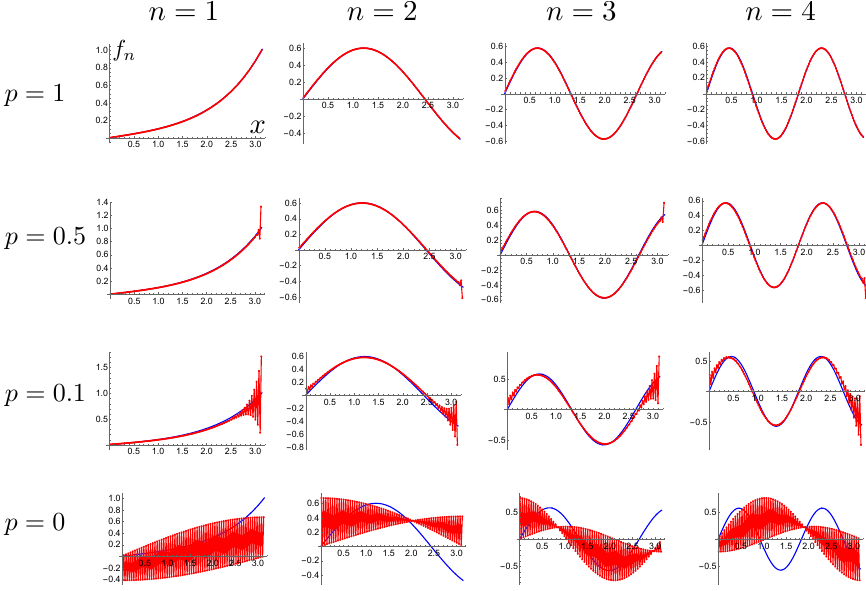}
\end{center}
\caption{Mode functions of QFT and spin systems for $p = 1, 0.5, 0.1,$ and $0$. 
The parameters are set to $L = 128$, $\ell = \pi$, and $m = 1$.
}
\label{mode_f}
\end{figure}

In Fig.~\ref{mode_f}, we compare mode functions of the QFT and the spin system.\footnote{As shown in Appendix~\ref{symmetry}, due to the spatial reflection symmetry of the theory, the mode functions $f_n$ and $g_n$ are related by spatial reflection as $g_n(x) = (-1)^{n+1} f_n(\ell - x)$. Therefore, only $f_n$ is shown here. In the case of $p = 0$, where degeneracy occurs, the eigenvectors are appropriately combined so that they become eigenvectors of the spatial reflection operator $P$. } 
The parameters are the same as those used in Fig.~\ref{omega}. We set $j_0=1/2$ in Eq.~(\ref{xjdef}).
The mode functions for $n = 1, 2, 3, 4$ are shown for $p = 1, 0.5, 0.1, 0$.  
The blue curve represents the mode function of the QFT, while the red dots and the connecting line show the results from the spin system. For $p = 1$, the mode functions of the two systems agree well.  
As $p$ deviates from unity, oscillatory modulations gradually appear in the mode functions.
The function $f_n$ exhibits such modulations near the boundary at $x=\ell$,
while $g_n$ shows the corresponding behavior near $x=0$. This behavior reflects the fact that the spin system increasingly deviates from
the boundary condition required to reproduce the continuum QFT,
Eq.~\eqref{pbc}:
the farther $p$ is from $1$, the less accurately the lattice boundary condition
approximates the smooth field-theoretical one.
Consequently, boundary-induced lattice-scale modulations become more pronounced.
For $p=0$, the low-energy sector is increasingly influenced by the coupling to the lattice doubler discussed above, leading to further deviations from the
continuum QFT, which extend into the bulk.

In Appendix~\ref{mode_near_boundary}, we analyze the behavior of the mode functions near the boundary in the spin system. As a result, we find that for $p \neq 1$, a mode exhibiting pronounced site-to-site oscillations emerges at the boundary. This behavior is consistent with the numerical results obtained here.


\subsection{Linear response}
\begin{figure}
\begin{center}
\includegraphics[scale=0.70]{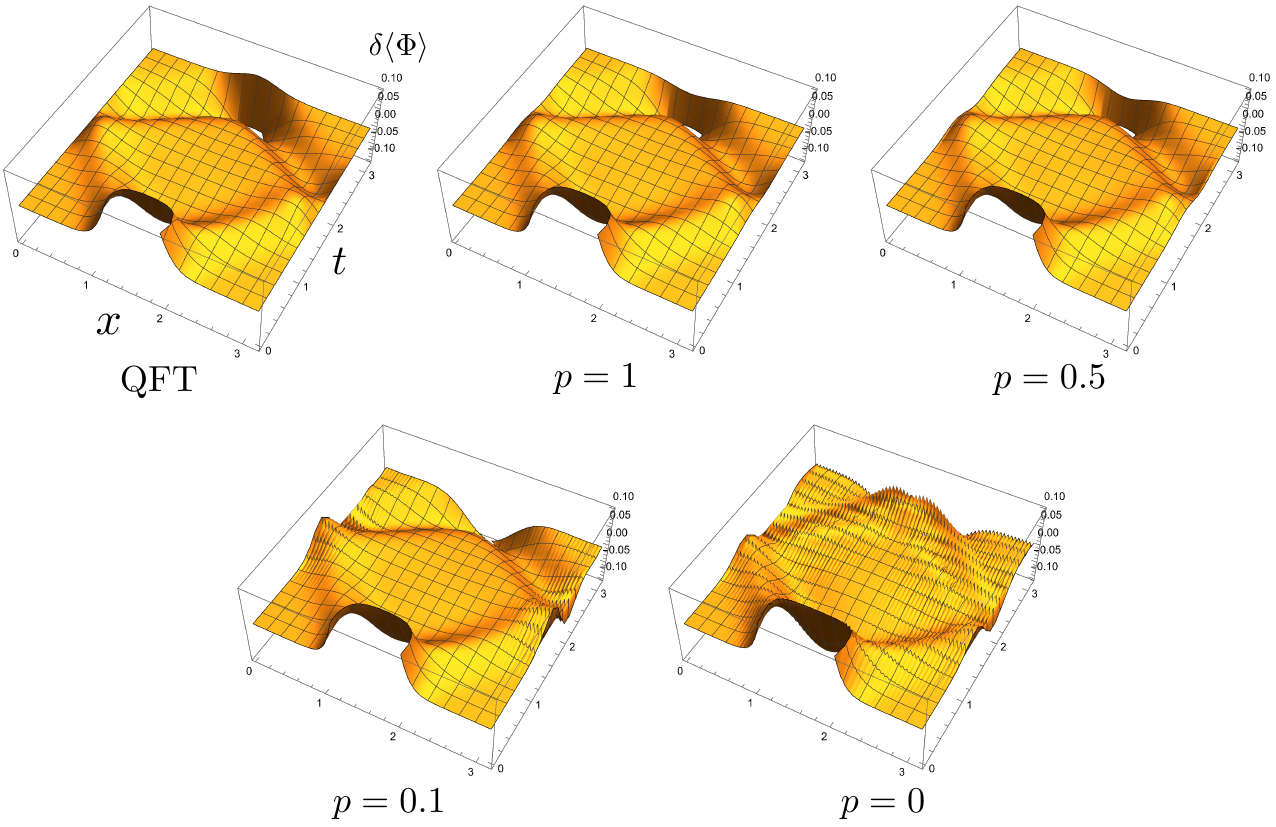}
\end{center}
\caption{Linear response of QFT and spin systems under a Gaussian source for $p = 1, 0.5, 0.1,$ and $0$. 
The parameters are set to $L = 128$, $\ell = \pi$, and $m = 1$.
}
\label{LR}
\end{figure}

To investigate how this mode function affects observables, let us consider the linear response. In QFT, we define a scalar operator 
\begin{equation}
    \Phi(t,x)=\frac{i}{2}\nord{\bar{\psi}(t,x)\psi(t,x)}=i\nord{a(t,x)b(t,x)}. 
\end{equation}
We denote the normal ordering of an operator $\mathcal{O}$ by $\nord{\mathcal{O}}$, which arranges all creation operators $\gamma_k^\dagger$ to the left of all annihilation operators $\gamma_k$, taking into account the spin-statistics signs. For example, $\nord{\gamma_n \gamma_{n'}^\dagger} = -\gamma_{n'}^\dagger \gamma_n$. We introduce the following perturbation to the Hamiltonian using the operator $\Phi$ as
\begin{equation}
    \delta H(t) = - \int^\ell_0  dxJ(t,x) \Phi(t,x)\ .
\end{equation}
The linear response to this perturbation is given by
\begin{equation}
    \delta \langle \Phi(t,x) \rangle = -\int_{-\infty}^\infty dt' \int_0^\ell dx' G_R(t,x,t',x')J(t',x')\ ,
\end{equation}
Here, $\delta \langle \Phi \rangle$ represents the deviation of the expectation value of $\Phi$ from its vacuum expectation value.
The retarded Green function $G_R$ is defined by
\begin{equation}
    G_R(t,x,t',x')= -i \theta(t-t') \langle [\Phi(t,x),\Phi(t',x')] \rangle \ ,
\end{equation}
where $\langle \cdots \rangle$ denotes the vacuum expectation value and $\theta(\tau)$ is the Heaviside step function ($\theta(\tau)=1$ for $\tau\ge 0$ and $\theta(\tau)=0$ for $\tau<0$). Using Eq.~(\ref{absol}), we have the explicit expression for the regarded Green function as
\begin{multline}
    G_R(t,x,t',x')=2\theta(t-t')\textrm{Im}\bigg[
    \langle a(t,x) a(t',x')\rangle \langle b(t,x) b(t',x')\rangle\\
    -\langle a(t,x) b(t',x')\rangle \langle b(t,x) a(t',x')\rangle\bigg]\ ,
\end{multline}
where two point functions of $a$ and $b$ are given by
\begin{equation}
\begin{split}
    &\langle a(t,x) a(t',x')\rangle=\sum_{n=1}^\infty f_n(x)f_n(x')e^{-i\omega_n (t-t')}\ ,\\
    &\langle b(t,x) b(t',x')\rangle=\sum_{n=1}^\infty g_n(x)g_n(x')e^{-i\omega_n (t-t')}\ ,\\ 
    &\langle a(t,x) b(t',x')\rangle=i\sum_{n=1}^\infty f_n(x)g_n(x')e^{-i\omega_n (t-t')}\ ,\\
    &\langle b(t,x) a(t',x')\rangle=-i\sum_{n=1}^\infty g_n(x)f_n(x')e^{-i\omega_n (t-t')}\ .
    \end{split}
    \label{twoptfuncs}
\end{equation}
As the source $J(t,x)$, we adopt the gaussian profile as
\begin{equation}
    J(t,x)=\exp\left[-\frac{t^2}{2 \sigma_t^2} - \frac{(x - \pi/2)^2}{2\sigma_x^2}\right]\ .
    \label{Jexp}
\end{equation}
In the explicit calculation, we set $\sigma_t = \sigma_x = 0.2$. 
With this Gaussian source, the contributions to the linear response in momentum space are suppressed for $k_n \gtrsim 1/\sigma_t$. 
Therefore, we set the upper limit of the summation in Eq.~(\ref{twoptfuncs}) to $N_{\textrm{max}} = 16$. 
We have confirmed that varying this $N_{\textrm{max}}$ does not significantly affect the results.
The upper-left panel of Fig.~\ref{LR} shows the result of the linear response in field theory. 
For clarity, the vertical axis is restricted to the range $|\delta\langle \Phi \rangle |\leq 0.1$.

Next, we turn to the discrete theory. We consider the perturbation of the Hamiltonian as
\begin{equation}
    \delta H(t) = - \varepsilon \sum_{j=1}^L  J(t,x_j) \Phi(t,x_j)\ ,
\end{equation}
where the scalar operator $\Phi(t,x_j)$ is given in terms of operators in the spin system $a_j(t)$ and $b_j(t)$ as 
\begin{equation}
    \Phi(t,x_j) = \frac{i}{\varepsilon}\nord{a_j(t) b_j(t)}\ .
\end{equation}
The linear response to this perturbation is given by
\begin{equation}
    \delta \langle \Phi(t,x_j) \rangle = -\varepsilon\sum_{j'=1}^L \int_{-\infty}^\infty dt'  G_R(t,x_j,t',x_{j'})J(t',x_{j'})\ .
\end{equation}
The retarded Green function is 
\begin{equation}
\begin{split}
    G_R(t,x_j,t',x_{j'})&= -i \theta(t-t') \langle [\Phi(t,x_j),\Phi(t',x_{j'})] \rangle \\
    &=\frac{2}{\varepsilon^2}\theta(t-t')\textrm{Im}\bigg[
    \langle a_j(t) a_{j'}(t')\rangle \langle b_j(t) b_{j'}(t')\rangle\\
    &\hspace{5cm} -\langle a_j(t) b_{j'}(t')\rangle \langle b_j(t) a_{j'}(t')\rangle\bigg]\ .
    \end{split}
\end{equation}
Two point functions of $a_j$ and $b_j$ are given by
\begin{equation}
\begin{split}
    &\langle a_j(t) a_{j'}(t')\rangle=\sum_{n=1}^L f_{jn}f_{j'n}e^{-i\omega_n (t-t')}\ ,\\
    &\langle b_j(t) b_{j'}(t')\rangle=\sum_{n=1}^L g_{jn}g_{j'n}e^{-i\omega_n (t-t')}\ ,\\ 
    &\langle a_j(t) b_{j'}(t')\rangle=i\sum_{n=1}^L f_{jn}g_{j'n}e^{-i\omega_n (t-t')}\ ,\\
    &\langle b_j(t) a_{j'}(t')\rangle=-i\sum_{n=1}^L g_{jn}f_{j'n}e^{-i\omega_n (t-t')}\ .
    \end{split}
    \label{twoptfuncs}
\end{equation}
We adopt the same source as that used in the linear response of QFT as in Eq.~(\ref{Jexp}).
Figure~\ref{LR} illustrates the linear response calculated in the spin system. 
We again examine the cases $p = 1, 0.5, 0.1,$ and $0$. 
For $p=1$, the spin system reproduces the QFT result well. 
As $p$ deviates from unity, fine oscillations appear near the boundary. 
This behavior is consistent with the fact that the mode functions oscillate near the boundary when $p \neq 1$.
In the case of $p=0$, it would be interesting that, although the mode functions themselves exhibit strong oscillations, the linear response still shows qualitatively similar behavior to that of QFT. 
Especially, before the wave packet reaches the boundary, the system is expected to provide a good approximation to the linear response of QFT even when $p \neq 1$.

\subsection{Relation to the Kitaev chain model}
\label{kitaevchain}
At this point, it is useful to note that the Hamiltonian considered above, Eq.~\eqref{HamiltonianJW}, is closely related to the Kitaev chain model~\cite{Kitaev01}, which describes a one-dimensional spinless $p$-wave superconductor. The Kitaev chain Hamiltonian is given by
\begin{equation}
H_{\rm K}=\sum_{j}\left[- w (c_j^\dagger c_{j+1}+c_{j+1}^\dagger c_j)-\Delta c_{j+1}c_j-\Delta^\ast   c_j^\dagger c_{j+1}^\dagger -\mu\left(c_j^\dagger c_j-\frac{1}{2}\right) \right]. \label{HK}
\end{equation}
Comparing Eqs.~(\ref{HamiltonianJW}) and~(\ref{HK}), we identify the parameters as
\begin{equation}
w=\frac{p}{2\varepsilon}, \qquad
\Delta=\Delta^\ast=\frac{1}{2\varepsilon}, \qquad
\mu=-\frac{p-m\varepsilon}{\varepsilon}. \label{eq:parameter_Kitaev}
\end{equation}
Accordingly, in units of the superconducting pairing strength
$\Delta=1/2\varepsilon$, the parameter $p$ controls the hopping amplitude,
while the combination $p-m\varepsilon$ determines an effective chemical potential. 

The Kitaev chain exhibits two distinct phases depending on the ratio between
the chemical potential $\mu$ and the hopping amplitude $w$~\cite{Kitaev01}.
For $|\mu|<2|w|$, the system is in a topological superconducting phase,
characterized by the presence of Majorana zero modes localized at the ends
of an open chain.
In contrast, for $|\mu|>2|w|$, the system is in a topologically trivial phase,
in which no such edge modes appear.
Using the parameter identification obtained above, the condition for the
topological phase can be expressed in terms of the original parameters as 
\begin{equation}
    |p-m\varepsilon| < |p|\ .
    \label{topocond}
\end{equation}
(This condition can be rephrased in the terms of the doubler mass $m_\textrm{doubler}$ as follows. $m>0$ and $m_\textrm{doubler}=2p/\varepsilon - m >0$, if $p>0$. $m<0$ and $m_\textrm{doubler}=2p/\varepsilon - m<0$, if $p<0$.)
Therefore, for $p>m\varepsilon/2$ under the current setup with $m,\varepsilon>0$,
the system lies on the topological side of the Kitaev chain phase diagram. 
In the continuum limit $\varepsilon \to 0$, the system approaches the phase boundary $|\mu| = 2|w|$ for any value of $p$, as can be seen from Eq.~(\ref{topocond}).

In the parameter set used in Fig.~\ref{mode_f}, this threshold corresponds to
$p=m\varepsilon/2=\pi/256\simeq 0.01$.
Thus, the lowest-energy mode ($n=1$) shown in Fig.~\ref{mode_f} can be identified
as the Majorana zero mode for $p=1,\,0.5,$ and $0.1$.
For a finite system, this mode represents a fermionic state formed by the
hybridization of two Majorana zero modes localized near the two ends of the chain;
the $f_n$ and $g_n$ components are localized near opposite boundaries,
although the localization appears weak due to the proximity to the phase boundary.

The presence or absence of oscillatory behavior in Majorana mode functions has
also been discussed in the context of the Kitaev chain~\cite{PhysRevB.82.094504,PhysRevB.86.220506,Prada:2012iv,Rainis:2013zxa,Hegde:2014tna,Hegde:2016ryf}.
It is important to note, however, that a direct comparison requires some care.
In Refs.~\cite{Hegde:2014tna,Hegde:2016ryf}, the oscillatory region is given by the condition
\begin{equation}
\frac{\mu^2}{4w^2}+\frac{|\Delta|^2}{w^2}<1 ,
\end{equation}
which characterizes an oscillatory \emph{decay} of the Majorana wave function,
where an exponentially decaying envelope is multiplied by a finite-wave-number
oscillatory factor.
Using the parameter identification in our model, this condition can be rewritten as
\begin{equation}
(m\varepsilon)^2 - 2p(m\varepsilon) + 1 < 1 .
\end{equation}
For $p<1$, this inequality is never satisfied for any value of $m\varepsilon$,
implying that all parameter sets considered in our calculations belong to the
non-oscillatory region in the sense of Refs.~\cite{Hegde:2014tna,Hegde:2016ryf}.
Entering the oscillatory region defined there would require $p>1$ and
$p-\sqrt{p^2-1} < m\varepsilon < p+\sqrt{p^2-1}$.

Nevertheless, the oscillations observed in Fig.~\ref{mode_f} are of a different
nature.
Rather than an oscillatory decay behavior, we observe a lattice-scale
modulation of the decay profile itself, i.e., a nonmonotonic envelope, which
persists even in the absence of oscillatory decay in the sense of the circle of
oscillations defined in Ref.~\cite{Hegde:2016ryf}.
Our formulation provides a unified perspective by showing that the suppression
of such lattice-scale modulations is a consequence of imposing the conditions
required to reproduce a smooth continuum QFT.
In particular, this requirement is implemented through the boundary condition
derived in Eq.~\eqref{pbc}, which directly constrains the structure of Majorana
modes in the lattice model.
Importantly, this argument is not restricted to the uniform Kitaev chain, but
extends to systems with spatially and/or temporally dependent parameters.


\section{Case of spatially dependent $p(x)$}
\label{sec:nonuniform_p}

\begin{figure}
\begin{center}
\includegraphics[scale=0.6]{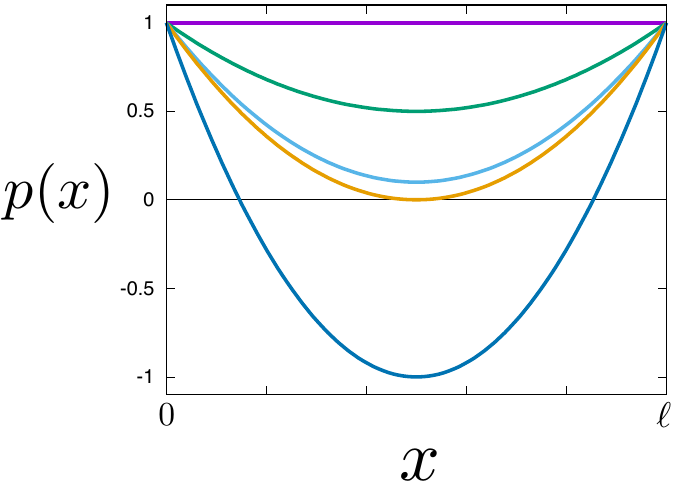}
\end{center}
\caption{Profiles of the function $p(x)$ for $\eta = 0, 0.5, 0.9, 1,$ and $2$ from top to bottom.
}
\label{p_of_x}
\end{figure}

As shown in Eq.~(\ref{pbc}), if the boundary values of the function $p(x)$ are properly set, the corresponding spin system is expected to approach QFT in the continuum limit. In this section, we consider the case where $p(x)$ depends on the spatial coordinate while setting the boundary value such that Eq.~(\ref{pbc}) is satisfied.
Here, as in the previous section, we consider flat spacetime and impose the same boundary conditions~(\ref{bc1}). As a specific form of the function $p(x)$, we take 
\begin{equation}
    p(x)=1+\eta\frac{4}{\ell^2}x(x-\ell)\ ,
\end{equation}
where $\eta$ is a constant parameter. Figure~\ref{p_of_x} shows the functional profile of the above function for $\eta=0,0.5,0.9,1,2$.
As discussed in Section~\ref{spec_and_modef}, when $p(x)$ is identically zero, the theory possesses doublers. Therefore, when $p(x)$ depends on space and crosses zero (i.e., for $\eta > 1$), one can expect local effects of the doubler to appear.
In this section, we analyze the effects of the spatial dependence of $p(x)$ on the spectrum and mode functions of the spin system, 
with particular attention to the influence of local doublers that appear when $p(x)$ crosses zero.

The Hamiltonian of the spin system is given by
\begin{equation}
\begin{split}
 H=&- \frac{1}{2\varepsilon} \bigg\{\sum_{j=1}^{L-1}\bigg[
  c_{j+1}c_j +c_j^\dagger c_{j+1}^\dagger
 + p_j( c_j^\dagger c_{j+1}+ c_{j+1}^\dagger c_j)
 +\sum_{j=1}^{L}q_j
(1-2c_j^\dagger c_j)\bigg] \bigg\}\\
=&\frac{1}{2}\pmat{\bm{c}^T & \bar{\bm{c}}^T}
\pmat{A & -S \\ S & -A} \pmat{\bm{c} \\ \bar{\bm{c}}}\ ,
 \label{IsingforQFTc2}
\end{split}
\end{equation}
where
\begin{equation}
    q_j\equiv p_j - \varepsilon\left(\frac{1}{2}p'_j + m\right)\ .
\end{equation}
Here, we use the shorthand notation: $p_j=p(x_j)$ and $p'_j=\partial_x p(x)|_{x=x_j}$.
We have also introduced the symmetric matrix $S$ and anti-symmetric matrix $A$ as
\begin{equation}
    S=-\frac{1}{2\varepsilon}\left(-2Q+P_++P_-\right)\ ,\quad 
    A=\frac{1}{2\varepsilon}(E_+-E_-)\ ,
\end{equation}
where $E_\pm$ are defined in Eq.~(\ref{Epmdef}). The matrices $Q$ and $P_\pm$ are defined by
\begin{equation}
Q=\begin{pmatrix}
q_1\, &  &   &    \\
  & q_2\, &  &    \\
  &   & \ddots &  \\
  &   &        & q_L\,
\end{pmatrix}
\ ,\quad
P_+=
    \begin{pmatrix}
0\, & p_1 &   &        &     \\
  & 0\, & p_2 &        &     \\
  &   & \ddots & \ddots &    \\
  &   &        & 0\, & p_{L-1} \\
  &   &        &   & 0\,
\end{pmatrix} \ ,\quad 
    P_-=P_+^T\ .
\end{equation}
By the same way as in Section~\ref{Dtheory}, we can obtain the spectrum and mode functions. For the general Hamiltonian~(\ref{IsingforQFTc}), one can similarly obtain a matrix representation. The corresponding expressions are summarized in Appendix.\ref{matrixform}.

Figure~\ref{omega_eta} shows the variation of the spectrum of the discrete theory as $\eta$ is varied as $\eta = 0, 0.5, 0.9, 1, 2$. The parameters are set to $L = 128$, $\ell = \pi$, and $m = 1$. For $\eta \gtrsim 1$, the difference between the spectra of the QFT and the discrete theory becomes larger than in the case of $\eta = 0$. This difference is likely due to the effect of local doublers. The local appearance of doublers can be readily confirmed by directly examining the mode functions.

\begin{figure}
\begin{center}
\includegraphics[scale=0.5]{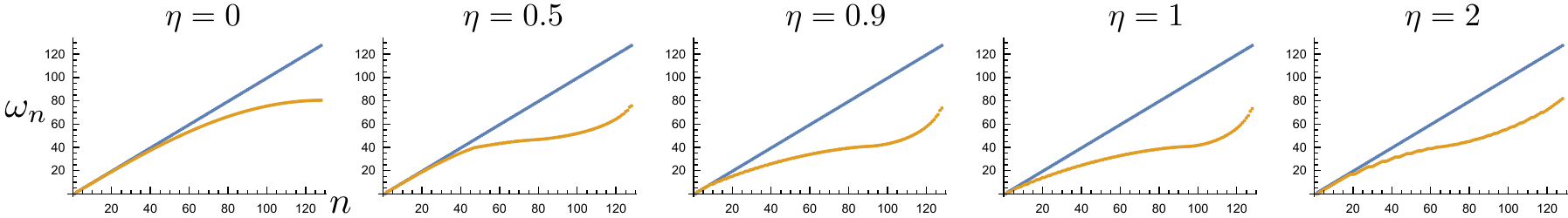}
\end{center}
\caption{Spectra of the QFT (blue dots) and spin systems (orange dots) for $\eta = 0, 0.5, 0.9, 1,$ and $2$. The parameters are set to $L = 128$, $\ell = \pi$, and $m = 1$.
}
\label{omega_eta}
\end{figure}

\begin{figure}
\begin{center}
\includegraphics[scale=0.5]{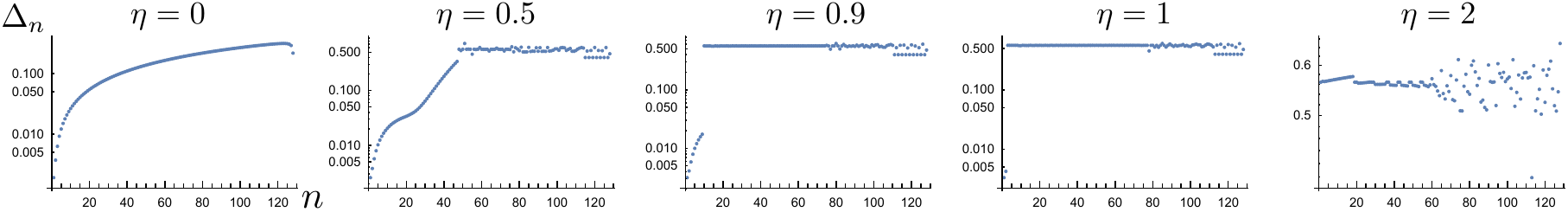}
\end{center}
\caption{Discretization error parameter $\Delta_n$ for $\eta = 0, 0.5, 0.9, 1,$ and $2$. 
The vertical axis is shown on a logarithmic scale. 
The parameters are set to $L = 128$, $\ell = \pi$, and $m = 1$.
}
\label{error}
\end{figure}

\begin{figure}
  \centering
\subfigure[$f_3(x)$ for $\eta=1$]
 {\includegraphics[scale=0.5]{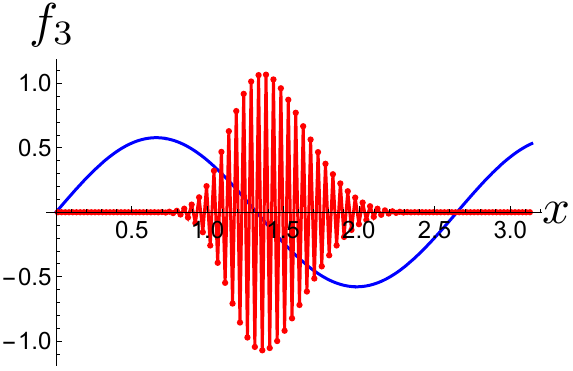}\label{f3eta1}
  }
  \subfigure[$f_1(x)$ for $\eta=2$]
 {\includegraphics[scale=0.5]{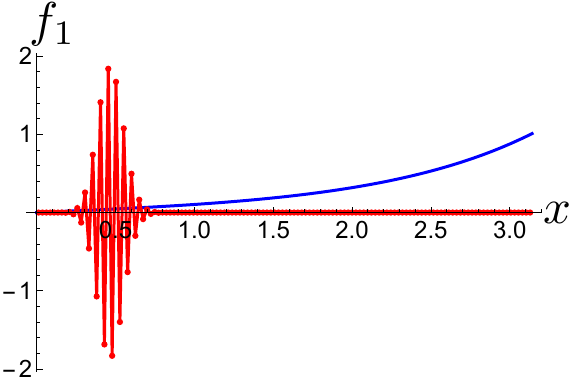}\label{f1eta2}
  }
 \caption{Mode functions $f_3(x)$ for $\eta = 1$ and $f_1(x)$ for $\eta = 2$.
 The parameters are set to $L = 128$, $\ell = \pi$, and $m = 1$.
}
 \label{f3f1}
\end{figure}

When $p(x)$ remains nonzero over the entire region, the mode functions are almost indistinguishable from those in the case where $p(x) = 1$. To quantify the difference between the mode functions $f_n(x)$ of the QFT and those of the discrete theory, we define the ``discretization error parameter'' $\Delta_n$ as 
\begin{equation}
    \Delta_n\equiv \sqrt{\frac{1}{L}\sum_{j=1}^L\left[f_{n}(x_j)-\frac{f_{jn}}{\sqrt{\varepsilon}}\right]^2}\ .
\end{equation}
Figure~\ref{error} shows the discretization error parameter $\Delta_n$ as a function of the mode number $n$. The vertical axis is plotted on a logarithmic scale. For $\eta \lesssim 1$, $\Delta_n$ takes small values in the region where $n \ll L$. 
(For example, when $\eta = 0$ or $0.5$, we find $\Delta_n \lesssim 0.1$ at $n/L \lesssim 0.3$. )
In contrast, for $\eta \gtrsim 1$, the error becomes large even in the region of small $n$.
For example, when $\eta = 1$, $\Delta_n$ already reaches $\sim 1$ at $n = 3$. 
Similarly, for $\eta = 2$, $\Delta_n$ becomes $\sim 1$ already at the lowest mode $n = 1$. 
What is happening to the mode functions in these cases? 
To examine this, Fig.~\ref{f3f1} shows the mode functions for $\eta = 1$, $n = 3$ and for $\eta = 2$, $n = 1$. From the figure, one can observe fine oscillations appearing around the region where $p(x) \sim 0$. This clearly indicates the excitation of a local doubler.

\begin{figure}
  \centering
\subfigure[$\eta=1$]
 {\includegraphics[scale=0.5]{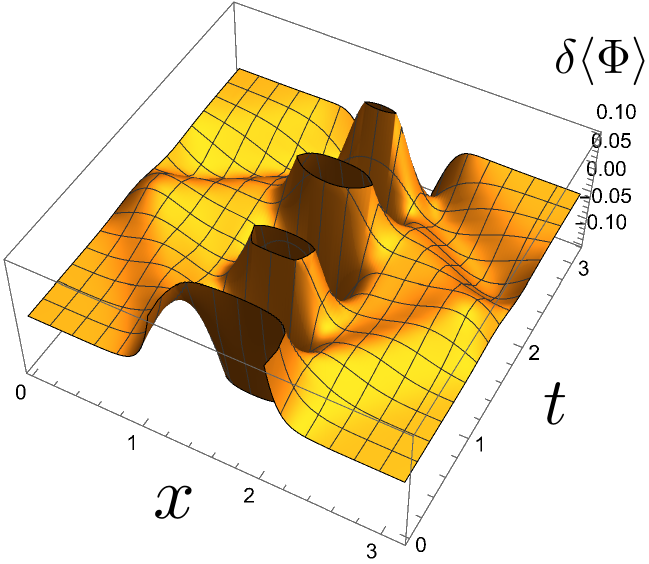}\label{LReta1}
  }
  \subfigure[$\eta=2$]
 {\includegraphics[scale=0.5]{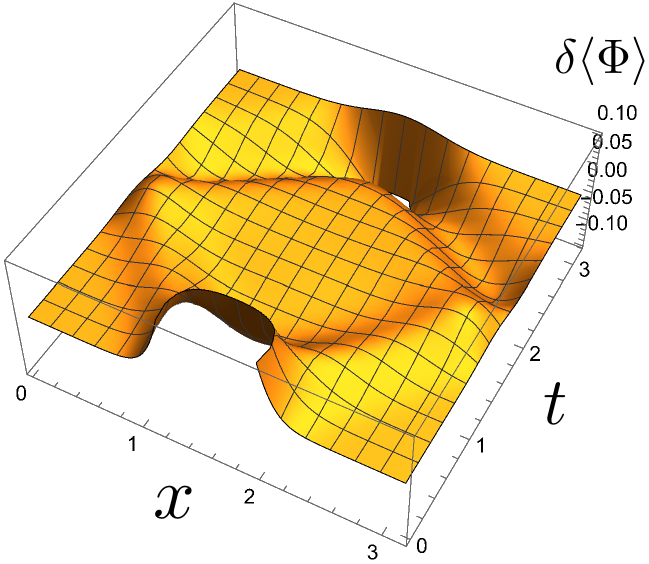}\label{LReta2}
  }
 \caption{Linear response of QFT and spin systems under a Gaussian source for $\eta = 1$ and $2$. The parameters are set to $L = 128$, $\ell = \pi$, and $m = 1$.
}
 \label{LR_eta1_2}
\end{figure}

As in the previous section, we analyze the linear response in order to investigate how local doublers affect physical observables. Figure~\ref{LR_eta1_2} shows the linear response to the Gaussian source given by Eq.~(\ref{Jexp}) for $\eta = 1$ and $2$. (The cases of $\eta = 0.5$ and $0.9$ were also computed, but since they show no significant difference from the $\eta = 0$ case, they are omitted here.) For $\eta = 1$, the response differs significantly from the QFT result. This is because, when $\eta = 1$, the local doubler is located around $x \sim 0$, where the Gaussian source is also applied, leading to strong excitation of the local doubler. In contrast, for $\eta = 2$, the result is almost identical to that of $\eta = 0$. This difference can be attributed to the fact that, unlike the $\eta = 1$ case, the position of the local doubler is displaced from that of the Gaussian source. The lesson obtained from this study is that while introducing spatial dependence in $p(x)$ is acceptable, one should, if possible, choose the configuration so as to avoid the appearance of local doublers.

\section{Summary}\label{sec:summary}

This paper extends the framework of using open quantum spin systems as quantum
simulators for quantum field theories (QFTs) with boundaries in two-dimensional
curved spacetimes.
Building on previous work that established a correspondence between spin systems
and QFTs with periodic spatial coordinates, we incorporate boundary conditions
and develop a systematic formulation applicable to systems with open boundaries,
which are ubiquitous in condensed-matter physics and quantum gravity.

We derive consistent boundary conditions for Majorana fermions by requiring the
preservation of the inner product in the continuum theory and show how these
conditions can be implemented on a discrete spin lattice.
This provides a concrete dictionary between boundary conditions in QFT and those
in the corresponding spin system.
As a benchmark, we analyze a flat-spacetime example and perform a detailed
comparison between the continuum QFT and the spin model, examining energy spectra,
mode functions, and linear response.
We find that, with appropriate parameter choices respecting the derived boundary
conditions, the spin system faithfully
reproduces the continuum QFT dynamics, including boundary-induced effects such as
edge modes, while deviations from these parameter choices lead to characteristic
lattice-scale modifications whose origin can be systematically understood.

As a direction for future research, we would like to consider realizing the moving mirror experiment in spin systems. In quantum field theory, it is known that an accelerating boundary gives rise to thermal radiation, which is often regarded as a simplified model of Hawking radiation. By performing an appropriate coordinate transformation, the spatial position of the moving boundary can be fixed, allowing us to apply the formalism developed in this paper. Furthermore, by employing the results of \cite{kinoshita2025spinsystemsquantumfield} together with an Unruh–DeWitt detector, we expect that the thermal properties of the moving mirror system can be probed through its excitation probability.

Beyond such direct applications to boundary QFT phenomena, the correspondence
established here also opens the possibility of reexamining known techniques in
quantum spin systems from a field-theoretical perspective.
One notable example is the sine-square deformation (SSD), a technique applied to quantum spin systems, in which the local energy scale is smoothly modulated according to a sine-squared function of position~\cite{Gendiar:2008udd,Hikihara:2011mtb,Katsura:2011zyx,Shibata:2011jup,Maruyama:2011njv,Katsura_2012}. This deformation suppresses boundary effects by gradually reducing the coupling strength toward the edges, effectively mimicking the behavior of an infinite or translationally invariant system. As a result, SSD has been shown to reproduce bulk properties with high accuracy even in finite-size systems, and it is widely used as a tool to study critical phenomena and conformal field theory aspects of quantum spin chains. Within our framework, it may be possible to elucidate the origin of the phenomena arising in SSD from a field-theoretical perspective.

\begin{acknowledgments}
The work of K.\ M.\ was supported in part by JSPS KAKENHI Grant Nos. JP21H05186 and JP22H01217.
The work of D.\ Y.\ was supported by JSPS KAKENHI Grant Nos.~JP21H05185, JP23K25830, and JP24K06890, and JST PRESTO Grant No.~JPMJPR245D. 
The work of R.\ Y.\ was supported by JSPS KAKENHI Nos.~JP19K14616, JP20H01838, and JP25K07156. 
\end{acknowledgments}

\appendix 

\section{Majorana fermions in the flat spacetime with boundaries}
\label{Majoranaflat}

We expand the time dependence of the Majorana fields as
\begin{equation}
\begin{split}
    a(t,x) &= \int_0^\infty d\omega \left( a_\omega(x)\, e^{-i\omega t} + a_{\omega}^\dagger(x)\, e^{i\omega t} \right)\ ,\\
    b(t,x) &= \int_0^\infty d\omega \left( b_\omega(x)\, e^{-i\omega t} + b_{\omega}^\dagger(x)\, e^{i\omega t} \right)\ ,
\end{split}
\end{equation}
so that the equation of motion for the Majorana field~(\ref{EOMab}) is written as
\begin{equation}
    \pmat{a_\omega \\ b_{\omega}}'
    =\pmat{m & i\omega \\ i\omega & -m}\pmat{a_\omega \\ b_{\omega}}\ .
    \label{abeq}
\end{equation}
By diagonalizing the above matrix, the differential equation can be solved.
In the following, we solve this equation separately for three cases: $\omega > m$, $\omega < m$ and $\omega=m$.

\subsection{Bulk modes: $\omega > m$}

We first consider the case $\omega > m$. In this case, the mode functions extend over the entire bulk region, and we shall refer to such modes as \textit{bulk modes}. In this case, the matrix in the equations of motion is diagnalized as
\begin{equation}
    P_k^{-1}\pmat{m & i\omega \\ i\omega & -m}P_k=\pmat{ik & 0\\ 0 & -ik}\ ,
\end{equation}
where 
\begin{equation}
k=\sqrt{\omega^2-m^2}\ ,\quad 
    P_k= \pmat{w_{-k} & w_{k}\\ -i w_{k} & -i w_{-k}}\ ,\quad
    w_{\pm k}=\frac{\omega-m\pm ik}{2\sqrt{\omega (\omega-m)}}\ .
\end{equation}
One can verify that $|w_k|^2 = |w_{-k}|^2 = w_k w_{-k} = 1/2$, and $w_{\pm k}^2 = (-m \pm ik)/(2\omega)$.
Thus, the solutions to the equations of motion are given by
\begin{equation}
    a_\omega(x)=w_{-k} e^{ikx} c_k +w_{k} e^{-ikx} d_k\ ,\quad 
    b_\omega(x) =-i(w_{k} e^{ikx} c_k+w_{-k} e^{-ikx} d_k)\ ,
    \label{aomegabomega}
\end{equation}
where $c_k$ and $d_k$ are constant operators. The boundary conditions~(\ref{bc1}) are given by $a_\omega|_{x=0} = b_\omega|_{x=\ell} = 0$. From the above solution, these boundary conditions can be written as
\begin{equation}
    \pmat{w_{-k} & w_k \\ w_{k} e^{ik\ell} & w_{-k} e^{-ik\ell}}
    \pmat{c_k \\ d_k} = 0\ .
\end{equation}
The existence of non-trivial solutions requires
\begin{equation}
    e^{2ik\ell} =\frac{w_{-k}^2}{w_k^2}=\frac{m+ik}{m-ik}\ .
\end{equation}
Therefore, the wave number $k$ is determined by
\begin{equation}
    \tan k\ell = \frac{1}{m\ell} k\ell \ .
\end{equation}
This equation has infinitely many solutions. We note that the solutions asymptotically behave as $k \sim \pi/2+\pi (n-1)$ for large $k$. We label the positive solutions in ascending order as $k = k_n$ ($n = 2, 3, \cdots$). (We reserve $n=1$ for the edge mode that will be introduced shortly below.) 
We also denote the corresponding eigenfrequencies as $\omega_n = \sqrt{k_n^2 + m^2}$. 
Under the assumption $\omega > m$, the solution with $k = 0$ does not need to be considered here. The case $\omega = m$ will be discussed separately in Section~\ref{omegam}.

From the boundary condition at $x=0$, we have $w_{-k} c_k+w_{k}d_k=0$. 
We introduce the operator $\gamma_n$ as
\begin{equation}
\pmat{c_k \\ d_k}\propto 
\pmat{w_{k} \gamma_n \\ -w_{-k} \gamma_n}\ .
\end{equation}
Then, we have 
\begin{equation}
\pmat{a_\omega(x) \\ b_\omega(x)}\propto
\pmat{ i \sin k_nx \\ (k_n\cos k_nx-m\sin k_nx)/\omega}\gamma_n\ .
\end{equation}
Overall coefficient will be determined by the orthonormal conditions of mode functions.

\subsection{Edge modes: $\omega < m$}
Next, we consider the case $\omega < m$. In this case, the mode functions are exponentially localized near the boundaries, and we shall refer to such a mode as an \textit{edge mode}.
In this case, the matrix in equations of motion is diagnalized as
\begin{equation}
    P_\kappa^{-1}\pmat{m & i\omega \\ i\omega & -m}P_\kappa=\pmat{\kappa & 0\\ 0 & -\kappa}\ ,
\end{equation}
where 
\begin{equation}
\kappa=\sqrt{m^2-\omega^2}\ ,\quad  
    P_\kappa= \pmat{w_{-\kappa} & w_{\kappa}\\ -i w_{\kappa} & -i w_{-\kappa}}
    \ ,\quad 
    w_{\pm \kappa}=\frac{\omega-m \pm \kappa}{2\sqrt{m(m-\omega)}}\ .
\end{equation}
One can verify $w_\kappa w_{-\kappa}=-\omega/(2m)$ and $w_{\pm\kappa}^2=(m\mp \kappa)/(2m)$. 
Thus, solutions of equations of motion are
\begin{equation}
    a_\omega(x)=w_{-\kappa}  e^{\kappa x} c_\kappa  +w_{\kappa } e^{-\kappa x} d_\kappa \ ,\quad 
    b_\omega(x) =-i(w_{\kappa } e^{\kappa x} c_\kappa +w_{-\kappa}  e^{-\kappa x} d_\kappa )\ ,
\end{equation}
where $c_\kappa$ and $d_\kappa$ are constant operators.
From boundary conditions~(\ref{bc1}), we have 
\begin{equation}
    \pmat{w_{-\kappa} & w_{\kappa} \\ w_{\kappa} e^{\kappa\ell} & w_{-\kappa} e^{-\kappa\ell}}
    \pmat{c_\kappa \\ d_\kappa} = 0\ .
\end{equation}
The existence of non-trivial solutions requires
\begin{equation}
    e^{2\kappa \ell}=\frac{w_{-\kappa}^2}{w_\kappa^2}=\frac{m+\kappa}{m-\kappa}\ .
\end{equation}
Therefore, we get
\begin{equation}
    \tanh \kappa\ell = \frac{1}{m\ell} \kappa \ell\ .
\end{equation}
For $m\ell > 1$, the above equation has a non-trivial solution. 
We assign $n = 1$ to the edge mode and denote its eigenfrequency as $\omega_1 = \sqrt{m^2 - \kappa^2}$. 
Under the assumption $\omega < m$, the solution with $\kappa = 0$ does not need to be considered here. The case $\omega = m$ will be discussed separately in next subsection.

From the boundary condition at $x=0$, we have $w_{-\kappa} c_\kappa+w_{\kappa}d_\kappa=0$. 
We introduce the operator $\gamma_1$ so that this equation is satisfied as
\begin{equation}
\pmat{c_\kappa \\ d_\kappa}\propto 
\pmat{-w_{\kappa} \\ w_{-\kappa}}\gamma_1 \ .
\end{equation}
Then, we obtain
\begin{equation}
\pmat{a_\omega(x) \\ b_\omega(x)}\propto
\pmat{ \frac{\omega}{m}\sinh \kappa x \\ \frac{i}{m}(m \cosh \kappa x-\kappa \sinh \kappa x)}\gamma_1\ .
\end{equation}
Overall coefficient will be determined by the orthonormal conditions of mode functions.

\subsection{Linear mode: $\omega=m$}
\label{omegam}
In this case, Eq.~(\ref{abeq}) becomes
\begin{equation}
    \pmat{a_\omega \\ b_{\omega}}'
    =m \pmat{ 1& i  \\ i  & -1}\pmat{a_\omega \\ b_{\omega}}\ .
\end{equation}
Its solution is
\begin{equation}
    \pmat{a_\omega \\ b_{\omega}}
    =\exp\bigg[ m\pmat{ 1& i  \\ i  & -1}x\bigg] \pmat{a_\omega(0) \\ b_{\omega}(0)}
    =\pmat{ 1+mx& imx  \\ imx  & 1-mx}\pmat{a_\omega(0) \\ b_{\omega}(0)}\ .
\end{equation}
Boundary conditions $a_\omega(0) = b_{\omega}(\ell) = 0$ are not satisfied in general, but it is fulfilled only in the special case where $m\ell = 1$. Then, the solution is given by
\begin{equation}
\pmat{a_\omega(x) \\ b_\omega(x)}\propto
\pmat{ ix/\ell \\ 1-x/\ell}\gamma_1\ ,
\end{equation}
where $\gamma_1$ is a constant operator. 
Since this mode does not coexist with the edge mode, we assign $n = 1$ to this mode as well.
We will refer this mode as the \textit{linear mode}.

\subsection{Mode functions}

In summary, the solution of the equation of motion for the Majorana field is given by
\begin{equation}
\begin{split}
&a(t,x)=i \sum_{n=1}^\infty f_n (x) (e^{-i\omega_n t}\gamma_n - e^{i\omega_n t}\gamma_n^\dagger)\ ,\\
&b(t,x)=\sum_{n=1}^\infty g_n(x) (e^{-i\omega_n t}\gamma_n + e^{i\omega_n t}\gamma_n^\dagger)\ ,
\end{split}
\label{absol0}
\end{equation}
where $f_n(x)$ and $g_n(x)$ are mode functions. For $n\geq 2$, they are given by
\begin{equation}
\begin{split}
    &f_n(x)=\left(\ell - \frac{m}{\omega_n^2}\right)^{-1/2}\sin k_n x\ ,\\
    &g_n(x)=\left(\ell - \frac{m}{\omega_n^2}\right)^{-1/2}
    \frac{1}{\omega_n}(k_n\cos k_n x-m \sin k_n x)
    \ .
\end{split}
\end{equation}
When $m\ell < 1$, neither the edge mode nor the linear mode exists. In this case, the summation in Eq.~(\ref{absol0}) is understood to start from $n = 2$.
For $m\ell > 1$, the edge mode is assigned to $n = 1$, and its mode function is given by
\begin{equation}
\begin{split}
    &f_1(x)=\left\{\frac{1}{m}\left(1-\frac{\ell \omega_1^2}{m}\right)\right\}^{-1/2}\frac{\omega_1}{m} \sinh\kappa x\ ,\\
    &g_1(x)=\left\{\frac{1}{m}\left(1-\frac{\ell \omega_1^2}{m}\right)\right\}^{-1/2}\frac{1}{m}(\kappa \cosh \kappa x-m\sinh \kappa x)\ .
\end{split}
\end{equation}
For $m\ell = 1$, the linear mode is assigned to $n = 1$, and its mode function is given by 
\begin{equation}
\begin{split}
    &f_1(x)=\sqrt{\frac{3}{2\ell}} \frac{x}{\ell}\ ,\\
    &g_1(x)=\sqrt{\frac{3}{2\ell}} \left(1-\frac{x}{\ell}\right)\ .
\end{split}
\label{linearmode}
\end{equation}
The coefficients of the mode functions are determined so that the following orthonormality condition holds:
\begin{equation}
\begin{split}
&\int^\ell_0 dx f_n f_{n'} =  \int^\ell_0 dx g_n g_{n'}=\frac{1}{2}\delta_{nn'}\ .
\end{split}
\end{equation}

\subsection{Majorana zero mode} 
We note that the condition of the existence of the zero mode $m=\kappa$ ($\omega_0=0$) becomes 
\begin{equation}
    \tanh \kappa\ell = 1 .
\end{equation}
This equation implies that the exact zero mode exists only when $\kappa \ell =\infty$. Thus, for example, the exact zero mode appears for the half-infinite system $\ell\to \infty$. 
Another possibility would be the heavy mass limit $\kappa=m=\infty$ with finite $\ell$. 

For both cases, the mode function for the edge mode becomes  
\begin{equation}
\begin{split}
    &f_1(x)=m e^{-m x}\ ,\\
    &g_1(x)=0\ , 
\end{split}
\label{zeromode1}
\end{equation}
and/or 
\begin{equation}
\begin{split}
    &f_1(x)=0\ ,\\
    &g_1(x)=m e^{-m (\ell-x)}\ , 
\end{split}
\label{zeromode2}
\end{equation}
where the former mode and the latter mode are, respectively, localized in the vicinity of the left edge and the right edge. 
For $m\to \infty$ in the finite system, $f_1(x)$ or $g_1(x)$ becomes the Dirac delta function; $f_1(x)=\delta (x)$ for the former case and $g_1(x)=\delta (\ell-x)$ for the latter case.

\subsection{On the other boundary condition: $a(0)=a(\ell)=0$}

For the alternative boundary condition $a(0) = a(\ell) = 0$, 
the spectrum and mode functions can be calculated in the same manner as in the previous discussion. 
For $\omega > m$, the condition for the wave number $k_n$ is obtained from Eq.~(\ref{aomegabomega}) as
\begin{equation}
    \textrm{det}\pmat{w_{-k} & w_k \\ w_{-k} e^{ik\ell} & w_{k} e^{-ik\ell}}=0\ .
\end{equation}
This leads to $e^{2ik_n \ell} = 1$, and the allowed wavenumbers are given by
\begin{equation}
    k_n=\frac{\pi n}{\ell}\quad (n=1,2,\ldots)\ .
\end{equation}
For this boundary condition, it is straightforward to confirm that neither edge modes nor linear modes exist.
The mode functions are given by
\begin{equation}
\begin{split}
    &f_n(x)=\frac{1}{\sqrt{\ell}}\sin k_n x\ ,\\
    &g_n(x)=\frac{1}{\sqrt{\ell}}\frac{1}{\omega_n}(k_n\cos k_n x-m \sin k_n x)
    \ .
\end{split}
\end{equation}

\section{Parity transformation}
\label{symmetry}
In this part, we explain the parity symmetry of the Hamiltonian.
In flat spacetime, the QFT of a Majorana field is invariant under the spatial reflection 
$x \to \ell - x$ and $\psi(t, \ell-x) \to -\Cliff^0 \psi(t, x)$. 
Using complex variables, this transformation can be written as $\Psi(t, \ell-x) \to i\,\Psi(t, x)$. 
In the corresponding spin system, the transformation is given by $c_{L-j+1} \to i\,c_j$. 
Indeed, it can be readily verified that the Hamiltonian~(\ref{HamiltonianJW}) remains invariant under this transformation. Using the real variables $\bm{a}$ and $\bm{b}$, this parity transformation can be written as 
\begin{equation}
    \pmat{\bm{a}\\\bm{b}}\to \pmat{0 & -J \\ J & 0}\pmat{\bm{a}\\\bm{b}}\ ,
\end{equation}
where $J$ is the $L\times L$ exchange matrix defined by
\begin{equation}
    J=\pmat{ &  &  \quad 1\\
     & \iddots &  \\   
     1 \quad &  &  
     }
     \ .
\end{equation}
Invariance of the Hamiltonian under the above transformation implies
\begin{equation}
    [M,P]=0\ ,\quad P=-i\pmat{0 & -J \\ J & 0}\ .
\end{equation}
That is, $M$ and $P$ can be simultaneously diagonalized. Since $P^2 = 1$, the eigenvectors of $M$ can be classified as parity even ($P=1$) or odd ($P=-1$) as
\begin{equation}
    P\pmat{i\bm{f}_j\\\bm{g}_j} = \pm \pmat{i\bm{f}_j\\\bm{g}_j} \quad \Leftrightarrow \quad
    \bm{g}_j = \pm J \bm{f}_j\ .
\end{equation}
Therefore, the mode functions $\bm{f}_j$ and $\bm{g}_j$ are related to each other by spatial reflection. When $M$ has degeneracy (that is, when $p = 0$), the degenerate eigenvectors can be classified according to the eigenvalues of $P$.

Of course, the same argument can be applied to QFT.
In fact, one can directly check the following equality: 
\begin{equation}
\begin{split}
    f_n(\ell-x)&=\left(\ell - \frac{m}{\omega_n^2}\right)^{-1/2}\sin k_n (\ell-x)\\
    &=\left(\ell - \frac{m}{\omega_n^2}\right)^{-1/2}\left(\sin k_n \ell\cos k_nx-\cos k_n \ell\sin k_nx\right)\\
    &= \pm \left(\ell - \frac{m}{\omega_n^2}\right)^{-1/2}\frac{1}{\sqrt{k_n^2+m^2}}\left(k_n\cos k_nx-m\sin k_nx\right)=\pm g_n(x),
\end{split}
\end{equation}
where we use $\cos k_n \ell=\pm m/\sqrt{k_n^2+m^2},\ \sin k_n \ell=\pm k_n/\sqrt{k_n^2+m^2}$ derived from the relation $\tan k_n \ell = k_n/m$. 
Noticing that $(n-1)\pi-\pi/2< k_n\le (n-1)\pi+\pi/2$, we can further conclude that $f_n(\ell-x)=(-1)^{n-1}g_n(x)$ for $n\ge 2$. 
Using a similar procedure, the relations for the edge mode \eqref{edgemodefn} and that for the linear mode \eqref{linearmode} can also be confirmed (namely $f_1(\ell-x)=g_1(x))$. 
Moreover, the zero modes \eqref{zeromode1} and \eqref{zeromode2} can be transformed into the parity-odd mode by choosing $f^{(-)}_1(x)=me^{-mx}/\sqrt{2},g^{(-)}_1(x)=-me^{-m(\ell-x)}/\sqrt{2}$ and the parity-even mode by $f^{(+)}_1(x)=me^{-mx}/\sqrt{2},g^{(+)}_1(x)=m e^{-m(\ell-x)}/\sqrt{2}$. 

\section{Matrix form of Hamiltonian}
\label{matrixform}

After the Jordan–Wigner transformation, the Hamiltonian can be written in the following form using a block-diagonal matrix as
\begin{equation}
    H=\frac{1}{2}\pmat{\bm{c}^T & \bar{\bm{c}}^T}
\pmat{A^\ast & -S^\ast \\ S & -A} \pmat{\bm{c} \\ \bar{\bm{c}}}\ ,
\end{equation}
where $\bm{c}=(c_1,\dots,c_L)^T$ and $\bar{\bm{c}}=(c_1^\dagger,\dots,c_L^\dagger)^T$. 
We have also introduced the Hermitian matrix $S$ and the antisymmetric matrix $A$ (both of size $L \times L$) as
\begin{equation}
    S=-\frac{1}{2\varepsilon}\left[-2Q + P+P^\dagger\right]\ ,\quad 
    A=\frac{1}{2\varepsilon}(Z-Z^T)\ ,
\end{equation}
where matrices $P,Z$ and $Q$ are defined by
\begin{equation}
\begin{split}
&P=
    \begin{pmatrix}
0\, & p_1+i\beta_1 &   &        &        \\
  & 0\, & p_2+i\beta_2 &        &        \\
  &   & \ddots & \ddots &        \\
  &   &        & 0\, & p_{L-1}+i\beta_{L-1} \\
  &   &        &   & 0\,
\end{pmatrix} \ ,\\
&Z=
    \begin{pmatrix}
0\, & \alpha_1e^{-i\zeta_1}/\gamma_1 &   &        &        \\
  & 0\, & \alpha_2e^{-i\zeta_2}/\gamma_2 &        &        \\
  &   & \ddots & \ddots &        \\
  &   &        & 0\, & \alpha_{L-1} e^{-i\zeta_{L-1}}/\gamma_{L-1} \\
  &   &        &   & 0\,
\end{pmatrix} \ ,\\
&Q=\textrm{diag}(q_1,\cdots,q_L)\ ,
\end{split}
\end{equation}
with 
\begin{equation}
    q_j=\frac{1}{2}[2p_j - \varepsilon(\partial_x p_j + 2m\alpha_j- \partial_t\zeta_j - \beta_j\partial_x\zeta_j)]\ .
\end{equation}
Using the real variables $\bm{a}$ and $\bm{b}$ defined by $\bm{c}=(\bm{b}-i\bm{a})/\sqrt{2}$, we obtain
\begin{equation}
    H=\frac{i}{2}\pmat{\bm{a}^T & \bm{b}^T}
\pmat{\textrm{Im}(S+A) & \textrm{Re}(S-A)  \\ -\textrm{Re}(S+A) & \textrm{Im}(S-A) } \pmat{\bm{a} \\ \bm{b}}\ .
\end{equation}

\section{Mode function analysis of the spin system near the boundary}
\label{mode_near_boundary}

As seen in Section~\ref{sec:Comparison}, when $p \neq 1$ the mode functions exhibit rapid oscillations near the boundary. 
In this appendix, we analyze the mode functions of the spin system in the vicinity of the boundary and show that this behavior can be reproduced.

Consider the following eigenvalue equation for the mode functions given in Eq.~(\ref{fgeq}):
\begin{equation}
    B^TB \bm{v}=\omega^2 \bm{v}\ ,
\end{equation}
where $\bm{v}=(v_1,\cdots,v_L)^T$ is the eigenvector.
By an explicit calculation, we obtain
\begin{multline}
    B^T B-\omega^2= \frac{1}{4\varepsilon^2}\Bigr[
(\mu-2\rho) E
-2\nu (E_++E_-)-\rho (E_+^2+E_-^2)\\
-(1-p)^2 P_1
-(1+p)^2 P_L
\Bigr],
\end{multline}
where $P_1=\text{diag}(1,0,\cdots,0)$ and $P_L=\text{diag}(0,\cdots,0,1)$. 
We also introduce $\mu = 4 (1 + (p - m\varepsilon)^2 - \varepsilon^2 \omega^2 )$, 
$\nu = 2p(p - m\varepsilon)$, and $\rho = 1 - p^2$ for notational simplicity. From the above expression, we obtain the following recurrence relation for $v_i$: 
\begin{equation}
\begin{split}
&\mu\, v_i
-2\nu\,(v_{i-1}+v_{i+1})
-\rho\,(v_{i-2}+2v_i+v_{i+2})
= 0
\qquad (3 \le i \le L-2)\ ,\\
&\bigl(\mu-2\rho-(1-p)^2\bigr) v_1
-2\nu\, v_2
-\rho\, v_3
= 0\ ,\\
&(\mu-2\rho)\, v_2
-2\nu\,(v_1+v_3)
-\rho\, v_4
= 0\ ,\\
&(\mu-2\rho)\, v_{L-1}
-2\nu\,(v_{L-2}+v_L)
-\rho\, v_{L-3}
= 0\ ,\\
&\bigl(\mu-2\rho-(1+p)^2\bigr) v_L
-2\nu\, v_{L-1}
-\rho\, v_{L-2}
= 0\ .
\end{split}
\end{equation}
By assuming $v_i \propto z^i$ and constructing the characteristic equation corresponding to the bulk recurrence relation, we obtain
\begin{equation}
    \mu 
-2\nu\,(z+z^{-1})
-\rho\,(z^2+z^{-2}+2)
= 0\ .
\end{equation}
By introducing $t = z + z^{-1}$, the above equation reduces to 
\begin{equation}
    \rho t^2 + 2\nu t - \mu = 0\ .
    \label{teq}
\end{equation}

Let us first consider the case $p \neq 1$, namely $\rho \neq 0$. Then, the solutions of the above quadratic equation are given by
\begin{equation}
    t=t_{\pm} 
= \frac{-\nu \pm \sqrt{\nu^2 + \rho\mu}}{\rho}\ .
\end{equation}
Characteristic roots can be obtained from $t$ as
\begin{equation}
    z = \frac{t \pm \sqrt{t^2 - 4}}{2} \ .
\end{equation}
Expanding $z$ to first order in $\varepsilon$, we obtain the following four solutions:
\begin{equation}
    z=1+\sqrt{m^2-\omega^2}\varepsilon\ ,\quad 1-\sqrt{m^2-\omega^2}\varepsilon\ ,\quad 
    -\frac{1-p}{1+p}(1+m\varepsilon)\ ,\quad -\frac{1+p}{1-p}(1-m\varepsilon)\ .
    \label{zsol}
\end{equation}
Here, when $m^2 < \omega^2$, we interpret $\sqrt{m^2 - \omega^2}$ as 
$i \sqrt{\omega^2 - m^2}$. Among the solutions above, the latter two do not approach 1 in the limit $\varepsilon \to 0$. 
They correspond to eigenfunctions that vary rapidly from site to site. 
This is the origin of the oscillatory behavior of the mode functions for $p \neq 1$ observed in Section~\ref{sec:Comparison}.

Next, let us consider the case $p = 1$ ($\rho = 0$). 
In this case, Eq.~(\ref{teq}) becomes a first-order equation, and consequently the characteristic root $z$ reduces to only the first two solutions in Eq.~(\ref{zsol}). 
These roots approach 1 in the limit $\varepsilon \to 0$ and correspond to smoothly varying mode functions. 
In particular, when $m^2 < \omega^2$, the solutions exhibit exponential behavior, whereas for $m^2 > \omega^2$, they show trigonometric behavior. 
This reproduces the behavior of the edge modes and bulk modes observed in Section~\ref{Conttheory}.

\section{Lattice fermion}
\label{app:lattice_fermion}

In this appendix, we summarize the key properties of the ($1+1$)-dimensional Majorana fermion from the perspective of lattice fermion theory.

In two dimensions, the chirality operator is defined as   
\begin{equation}
    \Cliff^5 \equiv - \Cliff^0 \Cliff^1 .
\end{equation}
This operator satisfies $\{\Cliff^5, \Cliff^i\} = 0$ and $(\Cliff^5)^2=E$.
Since the covariant derivative acting on spinor fields (\ref{eq:covariantD_spinor}) involves only $E$ and $\Cliff^5$, the Dirac operator 
$\slashed{\nabla}\equiv e^\mu_i \Cliff^i \nabla_\mu$ anticommutes with $\Cliff^5$, namely $\{\slashed{\nabla}, \Cliff^5\} = 0$.
Therefore, even in curved spacetimes, the mass of a spinor field can be identified from the structure of the field equation, where the mass term couples left- and right-handed components of the spinor field.
In the chiral representation, the gamma matrices are given by 
\begin{equation}
    U\Cliff^0U^{-1} = i\sigma^y =
    \begin{pmatrix} 
        0 & 1 \\
        -1 & 0
    \end{pmatrix}, \quad 
    U\Cliff^1U^{-1} = - \sigma^x =
    \begin{pmatrix} 
        0 & -1 \\
        -1 & 0
    \end{pmatrix}, 
\end{equation}
and     
\begin{equation}
    U\Cliff^5 U^{-1} = \sigma^z =
    \begin{pmatrix} 
        1 & 0 \\
        0 & -1
    \end{pmatrix}
    .
\end{equation}
Here, the matrix $U$ is defined as  
\begin{equation}
    U = \frac{1}{\sqrt{2}} \pmat{1 & 1 \\ -1 & 1} .
\end{equation}
With these definitions, the field equation (\ref{EOM}) can be explicitly written in the decomposed form  
\begin{equation}
    (i\sigma^y e_0^\mu - \sigma^x e_1^\mu)\nabla_\mu \chi = m\chi ,
\end{equation}
where we have defined  
\begin{equation}
    \chi = \pmat{\chi^+ \\ \chi^-} \equiv U\psi .
\end{equation}

In discrete theories, an analogous argument can be made.
We consider an inhomogeneous Kitaev chain Hamiltonian, whose parameters depend on each fermion site, as 
\begin{equation}
\mathcal{H}_j = - w_{j+\frac{1}{2}} c_j^\dagger c_{j+1} - w_{j+\frac{1}{2}}^\ast c_{j+1}^\dagger c_j - \Delta_{j+\frac{1}{2}} c_{j+1}c_{j}-\Delta^\ast_{j+\frac{1}{2}}  c_{j}^\dagger c_{j+1}^\dagger-\mu_j\left(c_j^\dagger c_j-\frac{1}{2}\right) .
\end{equation}
This corresponds to a Hamiltonian for a Majorana fermion on a two-dimensional curved spacetime in the continuum limit.
The Heisenberg equations for $c_j$ is 
\begin{equation}
    i \dot{c}_j = -\Delta^\ast_{j+\frac{1}{2}} c_{j+1}^\dagger + \Delta^\ast_{j-\frac{1}{2}} c_{j-1}^\dagger 
    - w_{j+\frac{1}{2}} c_{j+1} - w^\ast_{j-\frac{1}{2}} c_{j-1} - \mu_j c_j , \label{eq:Heisenberg_ci2}
\end{equation}
which is identical to Eq.~(\ref{eq:Heisenberg_ci}).
We introduce new real operators $\chi^\pm_j$ as 
\begin{equation}
    c_j \equiv \frac{1}{\sqrt{2}} e^{-i\zeta_j /2} \left(e^{-i\pi/4} \chi^+_j + e^{i\pi/4} \chi^-_j\right) ,
\end{equation}
which satisfy the following anti-commutation relations $\{\chi^+_i, \chi^+_j\}=\{\chi^-_i, \chi^-_j\}=\delta_{ij}$ and $\{\chi^+_i,\chi^-_j\} = 0$.
Substituting these into Eq.~(\ref{eq:Heisenberg_ci2}) and separating into real and imaginary parts yields 
\begin{multline}
    \dot{\chi}^+_j + |\Delta|_{j+\frac{1}{2}}\chi^+_{j+1} - |\Delta|_{j-\frac{1}{2}}\chi^+_{j-1} \\
    + \Im \left(w_{j+\frac{1}{2}}e^{-i(\zeta_{j+1}-\zeta_j)/2}\right) \chi^+_{j+1}
    - \Im \left(w_{j-\frac{1}{2}}e^{-i(\zeta_{j}-\zeta_{j-1})/2}\right) \chi^+_{j-1} \\
    = - \left(\mu_j + \frac{\dot{\zeta}_j}{2}\right) \chi^-_j 
    - \Re \left(w_{j+\frac{1}{2}}e^{-i(\zeta_{j+1}-\zeta_j)/2}\right) \chi^-_{j+1}
    - \Re \left(w_{j-\frac{1}{2}}e^{-i(\zeta_{j}-\zeta_{j-1})/2}\right) \chi^-_{j-1} ,
    \label{eq:chiral_eq1}
\end{multline}
and 
\begin{multline}
    \dot{\chi}^-_j - |\Delta|_{j+\frac{1}{2}}\chi^-_{j+1} + |\Delta|_{j-\frac{1}{2}}\chi^-_{j-1} \\
    + \Im \left(w_{j+\frac{1}{2}}e^{-i(\zeta_{j+1}-\zeta_j)/2}\right) \chi^-_{j+1}
    - \Im \left(w_{j-\frac{1}{2}}e^{-i(\zeta_{j}-\zeta_{j-1})/2}\right) \chi^-_{j-1} \\
    = \left(\mu_j + \frac{\dot{\zeta}_j}{2}\right) \chi^+_j 
    + \Re \left(w_{j+\frac{1}{2}}e^{-i(\zeta_{j+1}-\zeta_j)/2}\right) \chi^+_{j+1}
    + \Re \left(w_{j-\frac{1}{2}}e^{-i(\zeta_{j}-\zeta_{j-1})/2}\right) \chi^+_{j-1} ,
    \label{eq:chiral_eq2}
\end{multline}
where we have set $\zeta_j$ such that 
\begin{equation}
    |\Delta|_{j+\frac{1}{2}} = \Delta_{j+\frac{1}{2}} e^{-i \frac{\zeta_{j+1}+\zeta_{j}}{2}}
\end{equation}
is satisfied.
The left-hand sides of Eqs.~(\ref{eq:chiral_eq1}) and (\ref{eq:chiral_eq2}) indicate that $\chi^+$ and $\chi^-$ correspond to right-moving and left-moving operators, respectively, whereas the right-hand sides induce mixing between $\chi^+$ and $\chi^-$.
Introducing a two-component spinor at the $j$-th site, $\chi_j = (\chi^+_j , \chi^-_j )^T$, Eqs.~(\ref{eq:chiral_eq1}) and (\ref{eq:chiral_eq2}) can be recast into the following matrix form:  
\begin{multline}
    i\sigma^y \left[\dot{\chi}_j 
    + \Im \left(w_{j+\frac{1}{2}}e^{-i(\zeta_{j+1}-\zeta_j)/2}\right) \chi_{j+1}
    - \Im \left(w_{j-\frac{1}{2}}e^{-i(\zeta_{j}-\zeta_{j-1})/2}\right) \chi_{j-1} \right] \\
    - \sigma^x \left( |\Delta|_{j+\frac{1}{2}}\chi_{j+1} - |\Delta|_{j-\frac{1}{2}}\chi_{j-1} \right) \\  
    = \left(\mu_j + \frac{\dot{\zeta}_j}{2}\right) \chi_j 
    + \Re \left(w_{j+\frac{1}{2}}e^{-i(\zeta_{j+1}-\zeta_j)/2}\right) \chi_{j+1}
    + \Re \left(w_{j-\frac{1}{2}}e^{-i(\zeta_{j}-\zeta_{j-1})/2}\right) \chi_{j-1} ,
    \label{eq:chiral_eq3}
\end{multline}
where the coefficient matrices on the left-hand side anticommute with $\sigma^z$, reflecting chiral symmetry.
The right-hand side of the above equation can be decomposed as 
    \begin{multline}
    \left[\mu_j + \frac{\dot{\zeta}_j}{2} 
    + \Re \left(w_{j+\frac{1}{2}}e^{-i(\zeta_{j+1}-\zeta_j)/2}\right)
    + \Re \left(w_{j-\frac{1}{2}}e^{-i(\zeta_{j}-\zeta_{j-1})/2}\right)\right] \chi_{j} \\
    + \Re \left(w_{j+\frac{1}{2}}e^{-i(\zeta_{j+1}-\zeta_j)/2}\right) \chi_{j+1}
    + \Re \left(w_{j-\frac{1}{2}}e^{-i(\zeta_{j}-\zeta_{j-1})/2}\right) \chi_{j-1} \\
    - \left[\Re \left(w_{j+\frac{1}{2}}e^{-i(\zeta_{j+1}-\zeta_j)/2}\right)
    + \Re \left(w_{j-\frac{1}{2}}e^{-i(\zeta_{j}-\zeta_{j-1})/2}\right)\right]\chi_{j} .
\end{multline}
The terms in the second and third lines correspond to the so-called Wilson term~\cite{Wilson:1974sk}, which vanishes in the continuum limit.
Thus, in the continuum limit, the mixing term between the right-moving and left-moving fields 
contributes to the mass term of the Majorana fermion on curved spacetime as 
\begin{equation}
    m \alpha_j \approx 
    \mu_j + \frac{\dot{\zeta}_j}{2} 
    + \Re \left(w_{j+\frac{1}{2}}e^{-i(\zeta_{j+1}-\zeta_j)/2}\right)
    + \Re \left(w_{j-\frac{1}{2}}e^{-i(\zeta_{j}-\zeta_{j-1})/2}\right) .
\end{equation}
In general, the lapse function $\alpha_i$ and the mass $m$ are not uniquely separable. This reflects the restoration of conformal symmetry in the massless limit.
Once the mass of the Majorana field is fixed as a nonzero constant, the lapse function of the two-dimensional spacetime metric is determined.

It is well known that putting fermionic fields on a lattice leads to the appearance of additional fermionic modes, commonly referred to as doublers.
In the present case, a doubler mode can be identified through the transformation $\tilde c_j \equiv (-1)^j c_j$. 
Applying this transformation to Eqs.~(\ref{eq:chiral_eq1}) and (\ref{eq:chiral_eq2}) reverse the signs of several terms.
From the transformed equations, the effective mass of the doubler can be extracted as  
\begin{equation}
    m_\textrm{doubler} \alpha_j \approx 
    - \mu_j - \frac{\dot{\zeta}_j}{2} 
    + \Re \left(w_{j+\frac{1}{2}}e^{-i(\zeta_{j+1}-\zeta_j)/2}\right)
    + \Re \left(w_{j-\frac{1}{2}}e^{-i(\zeta_{j}-\zeta_{j-1})/2}\right) .
\end{equation}

Let us evaluate the above quantities in the flat spacetime example discussed in the main text.
Setting the parameters to be the same as in Eq.~(\ref{eq:parameter_Kitaev}),  
\begin{equation}
    w_{j+\frac{1}{2}}=\frac{p}{2\varepsilon} ,\quad \Delta_{j+\frac{1}{2}} = \frac{1}{2\varepsilon}\quad (\zeta_j=0),\quad
    \mu_j = - \frac{p-m\varepsilon}{\varepsilon}
    ,
\end{equation}
we obtain the lapse function $\alpha_j = 1$ for the Majorana mass $m$, and the mass of the doubler is given by 
\begin{equation}
    m_\textrm{doubler} = \frac{2p}{\varepsilon} - m .
\end{equation}
It turns out that, in the continuum limit $\varepsilon\to 0$, the mass of the doubler becomes very large.

In the case of a finite open chain with $L$ sites, the sites $j=0$ and $j=L+1$ are absent.
Therefore, for example at $j=L$, the equations of motion are equivalent to imposing the constraint: 
\begin{multline}
    \left[i\sigma^y 
    \Im \left(w_{L+\frac{1}{2}}e^{-i(\zeta_{L+1}-\zeta_L)/2}\right) 
    - \sigma^x  |\Delta|_{L+\frac{1}{2}}  
    - \Re \left(w_{L+\frac{1}{2}}e^{-i(\zeta_{L+1}-\zeta_L)/2}\right) \right] \chi_{L+1} = 0
\end{multline}
in Eq.~(\ref{eq:chiral_eq3}), as discussed in Section~\ref{sec:BCforSpinSystem}.
If the determinant of the coefficient matrix vanishes, namely,
\begin{equation}
\begin{aligned}
\begin{vmatrix}
    \Re \left(w_{L+\frac{1}{2}}e^{-i(\zeta_{L+1}-\zeta_L)/2}\right) &
    |\Delta|_{L+\frac{1}{2}}
    - \Im \left(w_{L+\frac{1}{2}}e^{-i(\zeta_{L+1}-\zeta_L)/2}\right) \\
    |\Delta|_{L+\frac{1}{2}}
    + \Im \left(w_{L+\frac{1}{2}}e^{-i(\zeta_{L+1}-\zeta_L)/2}\right) &
     \Re \left(w_{L+\frac{1}{2}}e^{-i(\zeta_{L+1}-\zeta_L)/2}\right)
\end{vmatrix}
\\
= |w_{L+\frac{1}{2}}|^2 - |\Delta_{L+\frac{1}{2}}|^2
    = 0 ,
    \end{aligned}
\end{equation}
then $\chi_{L+1}=(\chi^+_{L+1}, \chi^-_{L+1})^T$ satisfies the non-trivial relation 
\begin{multline}
    \sqrt{|\Delta|_{L+\frac{1}{2}}
    + \Im \left(w_{L+\frac{1}{2}}e^{-i(\zeta_{L+1}-\zeta_L)/2}\right)} \chi^+_{L+1} \\
    = -s
    \sqrt{|\Delta|_{L+\frac{1}{2}}
    - \Im \left(w_{L+\frac{1}{2}}e^{-i(\zeta_{L+1}-\zeta_L)/2}\right)} \chi^-_{L+1} ,
\end{multline}
where $s = \mathrm{sgn} \left(\Re (w_{L+\frac{1}{2}}e^{-i(\zeta_{L+1}-\zeta_L)/2})\right)$. 
This relation coincides with the admissible boundary condition in the corresponding continuum field theory.

Hence, when $|w_{j+\frac{1}{2}}|^2 - |\Delta_{j+\frac{1}{2}}|^2 = 0$, identical to (\ref{pbc}), is imposed at the ends of the lattice, the corresponding boundary condition for the Majorana field is automatically satisfied in the continuum limit, and the doubler mode decouples.
In contrast, if this condition is not imposed at the lattice ends, the constraint enforces $\chi^\pm_j=0$ at those endpoints, thereby inducing an excitation of doubler mode at the boundary.

\bibliography{refs}

\end{document}